\RequirePackage{tikz}

\documentclass[pdflatex, sn-standardnature,iicol]{sn-jnl}

\jyear{2023}%

\theoremstyle{thmstyleone}%
\newtheorem{theorem}{Theorem}
\newtheorem{lemma}[theorem]{Lemma}
\newtheorem{corollary}[theorem]{Corollary}

\theoremstyle{thmstyletwo}%
\newtheorem{example}{Example}%
\newtheorem{remark}{Remark}%

\theoremstyle{thmstylethree}%

\usepackage{amssymb}
\usepackage{mathtools}
\usepackage[nodollardollar, error]{onlyamsmath} 
\usepackage{microtype}
\usepackage{enumitem}
\usepackage{algorithm}
\usepackage{multirow}
\usepackage{booktabs}
\usepackage{colortbl}
\usepackage{academicons}
\definecolor{orcidlogocol}{HTML}{A6CE39}

\newcommand{\satset}{\mathrm{Sat}}
\newcommand{\paths}{\mathrm{paths}}
\newcommand{\post}{\mathrm{post}}
\newcommand{\strat}{f}
\newcommand{\variablefunc}{v}
\newcommand{\musatset}{\satset_{\mu}^\variablefunc}

\newcommand{\lfp}[2]{\mathsf{lfp}\,#1.#2}
\newcommand{\gfp}[2]{\mathsf{gfp}\,#1.#2}
\newcommand{\new}[1]{{#1}}

\usepackage{tikz}
\usetikzlibrary{arrows, automata, shapes}
\tikzset{auto, >= stealth}
\tikzset{every edge/.append style={thick, shorten >= 1pt}}
\tikzset{initial/.style={draw, thick, <-, shorten <=1pt}}
\tikzset{player0/.style = {draw, thick, shape=circle, minimum size=8mm}}
\tikzset{player1/.style = {draw, thick, shape=rectangle, minimum size=8mm}}

\tikzset{robust/.style={line width=.16ex,line join=round}}

\let\Box\relax
\DeclareMathOperator{\Box}{%
	\text{%
		\tikz[baseline]{%
    			\draw[robust] (0ex,-.1ex) -- (0ex, 1.4ex) -- (1.5ex, 1.4ex) -- (1.5ex, -.1ex) -- cycle;%
		}%
	}%
}

\DeclareMathOperator{\Boxdot}{%
	\text{%
		\tikz[baseline]{%
    			\draw[robust] (0ex, -.1ex) -- (0ex, 1.4ex) -- (1.5ex, 1.4ex) -- (1.5ex, -.1ex) -- cycle;%
	    		\fill (.75ex, .65ex) circle (.15ex);%
    		}%
	}%
}

\let\Diamond\relax
\DeclareMathOperator{\Diamond}{%
	\text{%
		\tikz[baseline]{%
			\draw[robust] (0ex,.6ex) -- (.95ex, 1.55ex) -- (1.9ex, .6ex) -- (.95ex, -.35ex) -- cycle;%
		}%
	}%
}

\DeclareMathOperator{\Diamonddot}{%
	\text{%
		\tikz[baseline]{%
			\draw[robust] (0ex,.6ex) -- (.95ex, 1.55ex) -- (1.9ex, .6ex) -- (.95ex, -.35ex) -- cycle;%
			\fill (.95ex, .6ex) circle (.15ex);%
		}%
	}%
}

\DeclareMathOperator{\X}{%
	\text{%
		\tikz[baseline]{%
    			\draw[robust] (.75ex, .65ex) circle (.75ex);%
    		}%
	}%
}

\DeclareMathOperator{\Xdot}{%
	\text{%
		\tikz[baseline]{%
    			\draw[robust] (.75ex, .65ex) circle (.75ex);%
	    		\fill (.75ex, .65ex) circle (.15ex);%
    		}%
	}%
}

\DeclareMathOperator{\U}{%
	\text{%
		\tikz[baseline]{%
			\node[inner sep=0pt, anchor=base, font=\bfseries] {U};%
		}%
	}%
}

\DeclareMathOperator{\Udot}{%
	\text{%
		\tikz[baseline]{%
			\node[inner sep=0pt, anchor=base, font=\bfseries] {U};%
			\fill (.1ex, .9ex) circle (.15ex);%
		}%
	}%
}

\DeclareMathOperator{\W}{%
	\text{%
		\tikz[baseline]{%
			\node[inner sep=0pt, anchor=base, font=\bfseries] {W};%
		}%
	}%
}

\DeclareMathOperator{\Wdot}{%
	\text{%
		\tikz[baseline]{%
			\node[inner sep=0pt, anchor=base, font=\bfseries] {W};%
			\fill (-.025ex, 0.27ex) circle (.15ex);%
		}%
	}%
}

\newcommand{\exampleend}{%
    \tikz{%
        \draw (0, 0) -| ++(1.5ex, 1.5ex);%
    }%
}

\newcommand{\ctlsem}{V_{\text{CTL}}}
\newcommand{\ctlstarsem}{V_{\text{{CTL*}}}}

\begin{document}

\title[Robust Computation Tree Logic]{Robust Computation Tree Logic}


\author*[1]{\fnm{Satya} \sur{Prakash Nayak}}\email{sanayak@mpi-sws.org}

\author*[2,3]{\fnm{Daniel} \sur{Neider}}\email{daniel.neider@tu-dortmund.de}

\author*[1]{\fnm{Rajarshi} \sur{Roy}}\email{rajarshi@mpi-sws.org}

\author*[4]{\fnm{Martin} \sur{Zimmermann}}\email{mzi@cs.aau.dk}

\affil[1]{\orgname{Max Planck Institute for Software Systems}, \orgaddress{\city{Kaiserslautern}, \country{Germany}}}

\affil[2]{\orgname{TU Dortmund University}, \orgaddress{\city{Dortmund}, \country{Germany}}}

\affil[3]{\orgname{Center for Trustworthy Data Science and Security}, \orgaddress{\city{Dortmund}, \country{Germany}}}

\affil[4]{\orgname{Aalborg University}, \orgaddress{\city{Aalborg},  \country{Denmark}}}

\abstract{
It is widely accepted that every system should be robust in that ``small'' violations
of environment assumptions should lead to ``small'' violations of system guarantees, but it is less clear how to make this intuition mathematically precise.
While significant efforts have been devoted to providing notions of robustness for Linear Temporal Logic (LTL), branching-time logics, such as Computation Tree Logic (CTL) and CTL*, have received less attention in this regard. 
To address this shortcoming, we develop ``robust'' extensions of CTL and CTL*, which we name robust CTL (rCTL) and robust CTL* (rCTL*).
Both extensions are syntactically similar to their parent logics but employ multi-valued semantics to distinguish between ``large'' and ``small'' violations of the specification.
We show that the multi-valued semantics of rCTL make it more expressive than CTL, while rCTL* is as expressive as CTL*.
Moreover, we show that the model checking problem, the satisfiability problem, and the synthesis problem for rCTL and rCTL* have the same asymptotic complexity as their non-robust counterparts, implying that robustness can be added to branching-time logics for free.
}


\maketitle

\section{Introduction}
\label{sec:Intro}

Specifications for reactive systems are typically written as an implication~$\Phi\Rightarrow\Psi$ where $\Phi$ is an environment assumption and $\Psi$ is a system guarantee. 
However, the specification $\Phi\Rightarrow\Psi$ is even satisfied if the environment assumption $\Phi$ is violated, no matter how the system behaves. 
This behavior is clearly inadequate since the environment assumptions will inevitably be violated in the real world: the actual environment where the system will be deployed is often not entirely known at design time and, thus, can not be accurately and entirely formalized by the formula~$\Phi$.

There have been concentrated efforts in the literature to prevent reactive systems from behaving arbitrarily when the environment assumption is violated, typically by making the specifications robust to violations of the environment assumption.
For instance, Bloem et al.~\cite{BloemGHJ09}, Tarraf et al.~\cite{TDM08}, Doyen et al.~\cite{DoyenHLN10}, Ehlers et al.~\cite{EhlersT14}, and Tabuada et al.~\cite{TB6,TB9} have provided different ways of introducing robustness for specifications in Linear Temporal Logic (LTL).
All these approaches require additional assumptions or quantitative information from the designer, which is often tedious and hard to obtain.

This drawback has motivated Tabuada and Neider~\cite{TabuadaN16} to introduce a new logic, named robust LTL (rLTL), which provides robustness without relying on any additional assumptions or input from a designer beyond an LTL formula.
Among rLTL's main features are its ease of use (one simply ``dots'' temporal operators in existing LTL formulas) and the fact that adding robustness does not change the asymptotic complexity of the model checking, runtime monitoring, and synthesis problems~\cite{TabuadaN16,NEIDER2021104810,DBLP:journals/iandc/NeiderWZ22,DBLP:conf/cdc/AnevlavisPNT18,DBLP:conf/hybrid/AnevlavisNPT19,DBLP:journals/tocl/AnevlavisPNT22,DBLP:conf/hybrid/MascleNSTW020,Mascle_2022,Nayak22}.
\new{Inspired by this logic, there have been several works introducing robust extensions of different classes of temporal logics~\cite{NEIDER2021104810,DBLP:journals/iandc/NeiderWZ22,murano2023robust,Zimmermann23}.\footnote{A detailed discussion of these extensions and other related work is presented in Section~\ref{sec:related}.}} 

In this work, we investigate robust branching-time logics.
Such logics, like Computation Tree Logic (CTL) and CTL*, have received less attention in this regard.
A notable exception is the work of French et al.~\cite{FrenchMR07,MCCABEDANSTED2019126}, which introduces logics called RoCTL and RoCTL*.
However, this logic again uses operators that require a manual quantification of the violations of the environment assumptions.

To address this shortcoming, we develop robust extensions of CTL and CTL*, which we call robust CTL (rCTL) and robust CTL* (rCTL*), which are inspired by the notion of robustness in rLTL. 
Similar to rLTL, our new logics employ multi-valued semantics to track the degree of violations of a specification and are guided by two objectives. First, the syntax of rCTL and rCTL* is similar to CTL and CTL*, respectively.
Second, the notion of robustness in these logics is intrinsic rather than extrinsic, i.e., robustness does not rely on the designers to provide quantitative information about the specification, such as the number of violations permitted, ranks, cost, etc.

As a demonstration of how our notion of robustness works, consider a specification~$\Phi\Rightarrow\Psi$ for a robot deployed in an office-like environment.
The environment assumption $\Phi = \forall \Box \neg H$ states that the human workers in the office never visit the robot's dock.
On the other hand, the robot guarantee $\Psi = \forall\Box\exists\X R$ states: ``for all trajectories, regardless of the robot's current position, the robot can return to its dock in one time step'' (note that such a specification can not be expressed in LTL). 
Ideally, we would then want the following: 
\begin{itemize}
	\item if the office workers satisfy the assumption~$\Phi$, then the robot should also satisfy the guarantee~$\Psi$;
	\item if the office workers violate the assumption by visiting the dock a finite number of times before realizing their mistake and eventually not visiting it anymore, i.e., if they only satisfy $\forall \Diamond \Box \neg H$, then the robot should also satisfy $\forall\Diamond\Box\exists\X R$, i.e., the robot eventually should be able to return to its dock from any point; and
	\item if the office workers violate the assumption by visiting the dock infinitely often (or eventually always), i.e., if they satisfy $\forall\Box\Diamond \neg H$ (or $\forall\Diamond \neg H$), then the robot should satisfy $\forall\Box\Diamond\exists\X R$ (or $\forall\Diamond\exists\X R$, respectively).
\end{itemize}
We later show that the semantics of rCTL and rCTL* indeed captures such a notion of robustness.

The first two contributions of the paper are robust variants of the logics CTL and CTL*, namely rCTL (in Section~\ref{sec:rCTL-motivation}) and rCTL* (in Section~\ref{sec:rctl-star}), respectively.
Their semantics rely on many-valued truth values that capture the various degrees of how a specification can be violated.

After having introduced rCTL and rCTL*, we study their expressive power and compare them to existing logics such as LTL, rLTL, CTL, and CTL* (in Subsections~\ref{sec:express rCTL} and \ref{subsec:rctlstarexpressiveness}). 
Our key results are that rCTL is more expressive than CTL, while rCTL* has the same expressive power as CTL*.

Next, we provide efficient model-checking algorithms for rCTL and rCTL* to demonstrate that both logics can be effectively used for verification.
We establish that the rCTL model checking problem is $\mathrm{PTIME}$-complete (in Subsection~\ref{sec: rCTL MC}) and that the rCTL* model checking problem is $\mathrm{PSPACE}$-complete (in Subsection~\ref{sec:rctlstarmc}).
Note that this is the same asymptotic complexity as CTL and CTL* model checking, respectively.
Moreover, we show that the satisfiability and reactive synthesis problems for rCTL (in Subsections~\ref{sec:rctlsat} and \ref{sec:rctlsynthesis}) and rCTL* (in Subsections~\ref{sec:rctlstarsat} and \ref{sec:rctlstarsynthesis}) match the exact asymptotic complexity of their non-robust counterparts, i.e., $\mathrm{EXPTIME}$-completeness for rCTL and $\mathrm{2EXPTIME}$-completeness for rCTL*.
Thus, robustness can be added to branching-time logics ``for free''.
Table~\ref{table:results} shows an overview over our complexity results.

\begin{table*}
\centering
\caption{Summary of our results (in gray) and comparison to other logics. All problems are complete for the respective complexity class.}
\label{table:results}
\begin{tabular}{llll}\toprule
& Model Checking & Satisfiability & Synthesis \\
 \midrule
CTL & $\mathrm{PTIME}$ & $\mathrm{EXPTIME}$ & $\mathrm{EXPTIME}$ \\
\rowcolor{lightgray!50} rCTL & $\mathrm{PTIME}$ & $\mathrm{EXPTIME}$ & $\mathrm{EXPTIME}$ \\
LTL & $\mathrm{PSPACE}$ & $\mathrm{PSPACE}$ & $\mathrm{2EXPTIME}$\\
rLTL & $\mathrm{PSPACE}$ & $\mathrm{PSPACE}$ & $\mathrm{2EXPTIME}$\\
CTL* & $\mathrm{PSPACE}$ & $\mathrm{2EXPTIME}$ & $\mathrm{2EXPTIME}$ \\
\rowcolor{lightgray!50} rCTL* & $\mathrm{PSPACE}$ & $\mathrm{2EXPTIME}$ & $\mathrm{2EXPTIME}$ \\
\bottomrule
\end{tabular}
\end{table*}

This paper is an extension of a conference paper~\cite{DBLP:conf/nfm/NayakNRZ22}.
The new content includes all proofs missing from the conference paper, an example illustrating our rCTL model-checking procedure, more details about the embedding of rCTL and rCTL* into the modal $\mu$-calculus, and the investigation of the rCTL and rCTL* synthesis problems.

\section{Notation and Review of Computation Tree Logic}

In this section, we review the syntax and semantics of CTL, which expresses properties of Kripke structures. 

Throughout this paper, we fix a finite set $\mathcal{P}$ of atomic propositions.
A \textit{Kripke structure} $M = (S ,I,R,L)$ over $\mathcal{P}$ consists of
a set of states $S$, a set of initial states $I\subseteq S$, a transition relation $R\subseteq S\times S$ such that for all states $s$ there exists a state $s'$ satisfying $(s,s')\in R$, and a labeling function $L\colon S\rightarrow 2^{\mathcal{P}}$.
We say that $M$ is finite if it has finitely many states. In that case, we define the size of $M$ as $\lvert S\rvert$.

The set~$\post(s) = \{s'\in S \mid (s,s') \in R\}$ contains all successors of $s \in S$.
A path of the Kripke structure $M$ is an infinite sequence~$\pi = s_0s_1\cdots$ of states such that $s_{i+1} \in \post(s_i)$ for each $i\geq 0$. For a state $s$, let $\paths(s)$ denote the set of all paths starting from $s$.
Furthermore, for a path~$\pi$ and $i\geq 0$, let $\pi[i]$ denote the $i$-th state of $\pi$, and let $\pi[i..]$ denote the suffix of $\pi$ from index $i$ on.

\subsection{Syntax} 

CTL formulas are classified into state and path formulas. Intuitively, state formulas express properties of states, whereas path formulas express temporal properties of paths. For ease of notation, we denote state formulas and path formulas by Greek capital letters and Greek lowercase letters, respectively.
CTL state formulas over $\mathcal{P}$ are given by the grammar
\[\Phi \Coloneqq p \mid \Phi\vee\Phi \mid \Phi \wedge \Phi \mid \neg \Phi \mid \Phi\Rightarrow\Phi \mid \exists \varphi \mid \forall\varphi,\]
where $p\in \mathcal{P}$ and $\varphi$ is a path formula. CTL path formulas are given by the grammar
\[\varphi \Coloneqq \X \Phi \mid \Diamond \Phi  \mid \Box  \Phi \mid \Phi \U \Phi \mid \Phi\W \Phi,\]
where $\X,\Diamond,\Box,\U$, and $\W$ denote the operators next, eventually, always, until, and weak until, respectively.
Note that we include implication, conjunction (alternatively, disjunction), and weak until as part of the syntax, instead of derived operators. We do this to be consistent with the syntax of robust logics, where these operators can no longer be derived.
As we will see later, it is also instructive to include the operators eventually and always explicitly. 
\subsection{Semantics} 
\label{sec:CTL-semantics}

Slightly deviating from the usual approach, we define the CTL semantics using a mapping $\ctlsem$ that maps a state/path and a CTL formula to a truth value in $\mathbb{B} = \{0,1\}$. 
Also, some of our definitions are non-standard in order to be closer to the robust semantics introduced later.
However, let us stress that the definition below is equivalent to the usual semantics of CTL (see, e.g., Baier and Katoen~\cite{model_checking_book}).

Given a state $s$ and state formulas~$\Phi,\Psi$, CTL semantics is defined as follows:
\begin{align*}
    \ctlsem(s,p) & = \begin{cases}
        0 & \text{if }p\not \in L(s); \text{ and} \\
        1 & \text{if }p\in L(s), \\
    \end{cases}\\
    \ctlsem(s,\Phi\vee\Psi) & = \max \{\ctlsem(s,\Phi),\ctlsem(s,\Psi)\}, \\
    \ctlsem(s,\Phi\wedge\Psi) & = \min \{\ctlsem(s,\Phi),\ctlsem(s,\Psi)\}, \\  
    \ctlsem(s,\neg \Phi) & = 1-\ctlsem(s,\Phi), \\
     \ctlsem(s, \Phi \Rightarrow \Psi)& = \begin{cases}
         1 &\hspace{-.2cm} \text{if }\ctlsem(s, \Phi)\le\\
         &\hspace{-.2cm} \ctlsem(s, \Psi); \text{and} \\
         \ctlsem(s, \Psi)&\hspace{-.2cm}\text{otherwise, }\\
     \end{cases}\\
    \ctlsem(s,\exists \varphi) & = \max_{\pi \in \paths(s)} \ctlsem(\pi,\varphi), \\
    \ctlsem(s,\forall \varphi) & = \min_{\pi \in \paths(s)} \ctlsem(\pi,\varphi).
\end{align*}
Similarly, for a path $\pi$, the CTL semantics of path formulas is defined as given below:
\begin{align*}
    \ctlsem(\pi,\X \Phi) & = \ctlsem(\pi[1],\Phi), \\
    \ctlsem(\pi,\Diamond \Phi) & = \max_{i\geq 0} \ctlsem(\pi[i],\Phi), \\
    \ctlsem(\pi,\Box \Phi) & = \min_{i\geq 0} \ctlsem(\pi[i],\Phi), \\
    \ctlsem(\pi,\Phi \U \Psi) & = \begin{multlined}[t] \max_{j\geq 0} \min \{\ctlsem(\pi[j],\Psi), \\ \min_{0\leq i< j} \ctlsem(\pi[i],\Phi)\}, \end{multlined} \\
    \ctlsem(\pi,\Phi \W \Psi) & = \begin{multlined}[t]
        \min_{j\geq 0} \max \{\ctlsem(\pi[j],\Phi), \\ \max_{0\leq i\leq j} \ctlsem(\pi[i],\Psi)\}.
    \end{multlined}
\end{align*}

\section{Robust Computation Tree Logic}
\label{sec:rCTL-motivation}

In this section, we robustify CTL by generalizing the ideas underlying robust LTL to CTL, obtaining the logic rCTL. We describe the syntax and semantics of rCTL and discuss the relation and differences between rCTL and other temporal logics.

As discussed in the robot example in the introduction, we want to capture the notion of robustness in CTL by ensuring that a small violation in environment assumptions leads to a small violation of system guarantees.
To achieve that, we introduce a robust semantics for CTL. Following arguments given by Tabuada and Neider~\cite{TabuadaN16}, we first motivate the semantics of rCTL using an example. Consider the CTL path formula $\Box p$, where $p$ is an atomic proposition. The formula can be satisfied in only one way, namely when $p$ holds at every step, i.e., state, of the path. In contrast, the formula can be violated in several ways. Intuitively, $\Box p$ is violated in the worst manner when $p$ fails to hold at every step. Then, we would prefer a case where $p$ holds for finitely many steps. Even better would be the case when $p$ holds at infinitely many steps. Finally, among all possible ways $\Box p$ can be violated, we would prefer the situation where $p$ fails to hold for at most finitely many steps. Our robust semantics is designed to distinguish between satisfaction and these four different degrees of violation of $\Box p$. 
However, as convincing as this argument might be, a question persists: in which sense can we regard these
five alternatives as canonical?

We answer this question by interpreting the satisfaction of $\Box p$ as a counting problem. Recall that the semantics of $\Box p$ for a path $\pi$ is given by $\ctlsem(\pi,\Box p) = \min_{i\geq 0} \ctlsem(\pi[i],p)$.
Now, observe that the truth value of the CTL formula $\Box p$ for a path $\pi$ only depends on the number of occurrences of $0$'s and $1$'s in the infinite word $\alpha = \ctlsem(\pi[0],p)\ctlsem(\pi[1],p)\cdots \in \mathbb{B}^{\omega}$ but not on their order. From this perspective, $\Box p$ is violated in the worst manner when $p$ fails to hold at every step, which corresponds to the number of occurrences of $1$ in $\alpha$ being zero. The next degree of violation of $\Box p$ in which $p$ holds at finitely many steps corresponds to having a finite number of $1$'s. Similarly, the next degree of violation corresponds to having an infinite number of $1$'s and an infinite number of $0$'s. Among all the ways in which $\Box p$ is violated, the most preferred way corresponds to having finitely many $0$'s. Finally, the satisfaction of $\Box p$ corresponds to having zero $0$'s. Note that the position where $0$'s and $1$'s occur is irrelevant for our argument. Furthermore, note that by successively applying permutations that swap position $i$ with position $i + 1$ and leave all the remaining elements of $\mathbb{N}$ unaltered, one can transform any $\alpha \in \mathbb{B}^{\omega}$ into words of one of the following five forms: $1^{\omega},0^k1^{\omega},(01)^{\omega},1^k0^{\omega},0^{\omega}$.
It is not hard to verify that the five cases of violations of $\Box p$ that we discussed above amount to the words of the five forms given above. Thus, we conclude the need for five truth values to describe five different ways of counting $0$'s and $1$'s that correspond to five different canonical forms of violations of $\Box p$.

According to our motivating example $\Box p$, the desired semantics should have one truth value corresponding to true and four truth values corresponding to the different shades of false. For notational convenience, we denote these truth values by $b=(b_1,b_2,b_3,b_4)$ with $b_i\in \mathbb{B}$. Intuitively, for the formula~$\Box p$, $b_1$ captures whether $p$ holds at every step, $b_2$ captures whether $p$ fails to hold at most finitely many steps, $b_3$ captures whether $p$ holds at infinitely many steps, and $b_4$ captures whether $p$ holds at least once.
Note that these cases are monotonic, i.e., $b_i = 1$ implies $b_{i+1} = 1$. 
Hence, we obtain the set~$\mathbb{B}_4 = \{0000,0001,0011,0111,1111\}$ of truth values. The value $1111$ corresponds to true, and the others correspond to different shades of false as explained above. The truth values are ordered naturally as $0000< 0001< 0011< 0111< 1111$.

It remains to explain how the semantics of Boolean connectives are defined for these truth values.
The notion of a triangular-norm summarizes all the desirable properties of a many-valued conjunction (see P. H\'ajek~\cite{multi-valued-conjuction} for details), and it is natural to model conjunction and disjunction in $\mathbb{B}_4$ by min and max, respectively. Moreover, as in intuitionistic logic, we define the implication, denoted by $a\rightarrow b$ on the level of truth values, such that $c\leq a \rightarrow b$ if and only if $c \wedge a \leq b$ for every $c\in \mathbb{B}_4$. This leads to 
\[a\rightarrow b = \begin{cases}
1111 &\text{ if $a\leq b$; and}\\
b &\text{ otherwise.}
\end{cases}\]

However, the negation, denoted by $\overline{a}$ on the level of truth values, defined by $a\rightarrow 0000$ as in intuitionistic logic, is not compatible with our interpretation that all elements in $\mathbb{B}_4\setminus \{1111\}$ represent different shades of false and, thus, their negation should be $1111$.
\new{To make this point clear, we present in Table~\ref{Neg} the intuitionistic negation in $\mathbb{B}_4$ and the desired negation compatible with the interpretation of the truth values in $\mathbb{B}_4$.}
\begin{table}[h]
\centering
\caption{\new{Desired negation vs.\ intuitionistic negation in $\mathbb{B}_4$.}}\label{Neg}
\begin{tabular}{c@{\hskip 3em}cc}
\toprule
\textbf{Value} & \textbf{\begin{tabular}[c]{@{}c@{}}Desired \\ negation\end{tabular}} & \textbf{\begin{tabular}[c]{@{}c@{}}Intuitionistic \\ negation\end{tabular}} \\ \midrule
1111 & 0000 & 0000 \\
0111 & 1111 & 0000 \\
0011 & 1111 & 0000 \\
0001 & 1111 & 0000 \\
0000 & 1111 & 1111 \\ \bottomrule
\end{tabular}
\end{table}
\new{What is then the algebraic structure on $\mathbb{B}_4$ that supports the desired negation, dual to the intuitionistic negation? This very same problem was investigated in~\cite{da_costa}, and the answer is \emph{da~Costa} algebras.
Therefore, following the ideas introduced by rLTL and use \textit{da~Costa algebras} to define the negation (see Priest and Graham~\cite{da_costa} for details):} 
\[\overline{a} = \begin{dcases}
	        0000 & \text{if }a=1111;\text{ and} \\
	        1111 & \text{otherwise}.
	        \end{dcases}\]
In other words, ``true'' (1111) gets mapped to ``false'' (0000), while ``shades of false'' get mapped to ``true''.

It should be mentioned that working with a five-valued semantics has its price. As in intuitionistic logic, $\overline{\overline{a}}$ may not be equal to $a$ as evidenced by taking $a=0111$. Although it is still true that $\overline{\overline{a}} \rightarrow a$. Interestingly, we can think of double negation as quantization in the sense that true is mapped to true and all the shades of false are mapped to $0000$ (false). Hence,
double negation quantizes the five different truth values into two truth values (true and false) in a manner that is compatible with our interpretation of truth values.
\new{\begin{remark}
Although there are alternative ways to define negation that preserves its duality, i.e., $\overline{\overline{a}} = a$, our notion of negation (as in original rLTL paper~\cite{TabuadaN16}) has been proven useful in many applications (see, e.g., Anevlavis et al.~\cite{DBLP:journals/tocl/AnevlavisPNT22}).
\end{remark}}

\subsection{Syntax}
The syntax of rCTL matches that of CTL, save for dotting temporal operators for visual distinction.
Hence, formulas of rCTL are also classified into state and path formulas. 

rCTL state formulas over $\mathcal{P}$ are formed according to the grammar
\[\Phi \Coloneqq p \mid \Phi \vee \Phi \mid \Phi \wedge \Phi \mid \neg \Phi \mid \Phi \Rightarrow \Phi \mid \exists \varphi \mid \forall \varphi,\]
where $p\in \mathcal{P}$ and $\varphi$ is a path formula. rCTL path formulas are formed according to the grammar
\[\normalfont \varphi \Coloneqq \Xdot \Phi\mid \Diamonddot \Phi \mid \Boxdot \Phi \mid \Phi \Udot \Phi \mid \Phi \Wdot \Phi.\]

The size of a formula is defined as the number of its syntactically distinct subformulas. Here, the set of subformulas of a state formula~$\Phi$ is defined as for CTL (see Baier and Katoen~\cite{model_checking_book} for details) and denoted by $\mathrm{Sub}(\Phi)$.

\subsection{Semantics}
\label{sec:rCTL-semantics}

Similar to the semantics of CTL, we define the semantics of rCTL by a mapping $V$, called \textit{valuation}, that maps an rCTL formula and a state/path to an element of $\mathbb{B}_4$. 
For an atomic proposition $p\in \mathcal{P}$, it is defined classically:
\begin{align*}
	 V(s,p) & = \begin{dcases}
	        0000 & \text{if }p\not \in L(s);\text{ and} \\
	        1111 & \text{if }p\in L(s). \\
	    \end{dcases}
\end{align*}
Following the semantics of rLTL, we define the semantics for Boolean connectives in rCTL using da Costa algebras, as follows: 
\begin{align*}
    V(s,\Phi \vee \Psi) & = \max\{V(s,\Phi),V(s,\Psi)\}, \\
    V(s,\Phi \wedge \Psi) & = \min\{V(s,\Phi),V(s,\Psi)\}, \\
    V(s,\neg \Phi) & = \overline{V(s, \Phi)}, \\
    V(s,\Phi \Rightarrow \Psi) & = V(s,\Phi)\rightarrow V(s,\Psi).
\end{align*}
For existential path quantification, we want $V(s,\exists \varphi) \geq b$ if there exists a path~$\pi$ starting in $s$ such that $V(\pi,\varphi)\geq b$. Similarly, we want $V(s,\forall \varphi) \geq b$ if for all paths~$\pi$ starting in $s$ it holds that $V(\pi,\varphi)\geq b$. This leads to
\begin{align*}
    V(s,\exists \varphi) & = \max_{\pi \in \paths(s)} V(\pi,\varphi),\\
    V(s,\forall \varphi) & = \min_{\pi \in \paths(s)} V(\pi,\varphi).
\end{align*}
For path formulas, we formalize the intuition above in the semantics of the temporal operators. 
For $1\leq \ell \leq 4$, let $V_\ell$ denote the $\ell$-th bit of the valuation $V$.
Then, using the counting interpretation as discussed earlier, we define the semantics for $\Boxdot$ by $V(\pi,\Boxdot \Phi) = (b_1,b_2,b_3,b_4)$, where
\begin{align*}
    b_1 & = \min_{i\geq 0} V_1(\pi[i], \varphi),\\
    b_2 & = \max_{j\geq 0} \min_{i\geq j} V_2(\pi[i], \varphi),\\
    b_3 & = \min_{j\geq 0} \max_{i\geq j} V_3(\pi[i], \varphi),\\
    b_4 & = \max_{i\geq 0} V_4(\pi[i], \varphi).
\end{align*}

The semantics of $\Diamonddot\Phi$ mimics the classical semantics in that the truth value of $\Diamonddot\Phi$ on $\pi$ is the maximal truth value of $\Phi$ that is assumed at any position of~$\pi$.
Analogously, the semantics for temporal operators $\Xdot$ and $\Udot$ also mimics the classical semantics as follows:
\begin{align*}
    V(\pi,\Diamonddot \Phi) & = \max_{i\geq 0} V(\pi[i],\Phi),\\
    V(\pi,\Xdot \Phi) & = V(\pi[1], \Phi),\\
    V(\pi,\Phi\Udot \Psi) & = \begin{multlined}[t]
        \max_{j\geq 0} \min \{V(\pi[j],\Psi), \\
        \min_{0\leq i < j} V(\pi[i],\Phi)\}.
    \end{multlined}
\end{align*}

Finally, using the counting interpretation as above, the semantics for $\Wdot$ is defined by $V(\pi,\Phi\Wdot \Psi) = (b_1,b_2,b_3,b_4)$, where
\begin{align*}
    b_1 & = \min_{j\geq 0}\max \{V_1(\pi[j],\Phi), \max_{0\leq i \leq j} V_1(\pi[i],\Psi)\},\\
    b_2 & = \max_{k\geq 0} \min_{j\geq k}\max \{V_2(\pi[j],\Phi), \max_{0\leq i \leq j} V_2(\pi[i],\Psi)\},\\
    b_3 & = \min_{k\geq 0} \max_{j\geq k}\max \{V_3(\pi[j],\Phi), \max_{0\leq i \leq j} V_3(\pi[i],\Psi)\},\\
    b_4 & = \max_{j\geq 0}\max \{V_4(\pi[j],\Phi), \max_{0\leq i \leq j} V_4(\pi[i],\Psi)\}.
\end{align*}

\begin{example}\label{example_rctl}
Having defined the rCTL semantics, let us recall the example of the specification for a robot given in Section~\ref{sec:Intro}: $\forall\Box \neg H \Rightarrow \forall\Box \exists\X R$, where $\forall \Box \neg H$ is the environment assumption that human office workers
never visit the dock of the robot, and $\forall\Box\exists\X R$ is the robot guarantee that from every state in every path, i.e., from every reachable state, there exists a way for the robot to return to its dock in one time step. The robust version of this formula is $\Phi = \forall\Boxdot \neg H \Rightarrow \forall\Boxdot \exists\Xdot R$. Let us demonstrate how this formula captures the robustness property as discussed in Section~\ref{sec:Intro}.

Let us assume $\Phi$ evaluates to $1111$ in a given Kripke structure. Then the following hold:
\begin{itemize}
\item If the office workers never visit the dock, then in any path, $\neg H$ holds at every state. Hence, $\forall\Boxdot \neg H$ evaluates to $1111$. Then by the semantics of $\Rightarrow$, the formula $\forall\Boxdot \exists\Xdot R$ also must evaluate to $1111$. That means, in any path, $\exists\Xdot R$ also holds at every state. Therefore, from any state of a path, the robot can return to its dock in one time step. Hence, the desired behavior of the system is retained when the environment assumption holds with no violation.

\item If the office workers violate the assumption by visiting the dock finitely many times and eventually not visiting it anymore, then for any path, $\neg H$ holds eventually at every state. Hence, $\forall\Boxdot \neg H$ evaluates to $0111$. Then, by the rCTL semantics, $\forall\Boxdot \exists\Xdot R$ evaluates to $0111$ or higher. Hence, in any path, $\exists\Xdot R$ also needs to hold eventually at every state. That means, from any state in a path, the robot can return to its dock eventually. 

\item Similarly, if $\neg H$ holds at infinitely many states (some state) in every path, then $\exists\Xdot R$ needs to hold at infinitely many states (some state) in every path.
\end{itemize}
Hence, whenever the formula  $\Phi$ evaluates to $1111$, its semantics captures the intended robustness property by which a weakening of the assumption $\forall\Boxdot \neg H$ leads to a weakening of the guarantee $\forall\Boxdot\exists\Xdot R$.

Now, a natural question arises: does the formula still provide useful information when its value is lower than $1111$. It follows from the semantics of implication that $\Phi$ evaluates to $b<1111$ only when $\forall\Boxdot \neg H$ evaluates to a higher value than $b$, whereas $\forall\Boxdot \exists\Xdot R$ evaluates to $b$. So, the desired system guarantee is not satisfied. However, the value of $\Phi$ still describes which weakened guarantee follows from the environment assumption. This can be seen as another measure of robustness: despite $\forall\Boxdot \exists\Xdot R$ not following from
$\forall\Boxdot \neg H$, the system's behavior is not arbitrary, a value of $b$ is still guaranteed.
\hfill\exampleend
\end{example}

\new{It is worth mentioning that even though our notion of robustness is motivated by the robustness in formulas of the form $\Phi\Rightarrow\Psi$, such a notion has also value beyond this class of specifications. For example, the work of Anevlavis et al.~\cite{DBLP:journals/tocl/AnevlavisPNT22} shows that the relevant reactivity patterns~\cite{DwyerAC99} fall under the fragment of rLTL that does not contain the implication operator. }

\subsection{Expressiveness of rCTL}
\label{sec:express rCTL}

In this section, we compare the expressiveness of rCTL with three other temporal logics: CTL, LTL, and rLTL.
We show that the five truth values of rCTL make it more expressive than CTL.
More precisely, there are properties that one can express in rCTL but not in CTL.
However, the expressiveness of rCTL and LTL are incomparable, and the same also holds for rCTL and rLTL.

We compare the expressiveness of two classes of logics by comparing the expressiveness of their formulas. For logics $\mathcal{L}$ and $\mathcal{L}'$, we say $\mathcal{L}$ is as expressive as $\mathcal{L}'$ if for every formula in $\mathcal{L}'$ there is an equivalent formula in $\mathcal{L}$. Moreover, we say $\mathcal{L}$ is more expressive than $\mathcal{L}'$ if $\mathcal{L}$ is as expressive as $\mathcal{L}'$ but the converse is not true. Furthermore, we say $\mathcal{L}$ and $\mathcal{L}'$ have incomparable expressiveness if neither of $\mathcal{L}$ and $\mathcal{L}'$ is as expressive as the other one.

Now the question is what it means for two formulas to be equivalent. Intuitively speaking, equivalent means ``express the same thing''. Formally, we define the equivalence of two formulas using their satisfaction sets. For a given Kripke structure, and a state formula $\Phi$, we define the satisfaction set $\satset(\Phi,b)$ of an rCTL formula $\Phi$ and with value $b\in \mathbb{B}_4$ to be the set of states $s$ such that $V(s,\Phi)\geq b$. Since the satisfaction sets of an rCTL (state) formula are always associated with a truth value in $\mathbb{B}_4$, we always associate a truth value with an rCTL formula when comparing its expressiveness.

For two rCTL state formulas $\Phi_1, \Phi_2$ and two truth values $b_1,b_2\in \mathbb{B}_4$, we say that $\Phi_1$ with truth value $b_1$ is equivalent to $\Phi_2$ with truth value $b_2$ if for every Kripke structure it holds that $\satset(\Phi_1,b_1) = \satset(\Phi_2,b_2)$. 
Similarly, an rCTL formula~$\Phi_1$ with truth value $b_1$ is equivalent to a CTL formula~$\Phi_2$ if for every Kripke structure it holds that $\satset(\Phi_1,b_1) = \satset_{\text{CTL}}(\Phi_2)$, where $\satset_{\text{CTL}}(\cdot)$ denotes the satisfaction sets for CTL formulas. 

For an LTL (or rLTL) formula~$\varphi$ (which is evaluated over paths), we define its satisfaction set to contain all states~$s$ such that $\pi$ satisfies $\varphi$ for every path~$\pi \in \paths(s)$.
Hence, an LTL or rLTL formula is equivalent to an rCTL formula, if they have the same satisfaction sets for all Kripke structures.

We begin by comparing the semantics of CTL and rCTL. First, we want to show that the CTL semantics is captured by the first bit of the rCTL semantics (recall that $V_1$ denotes the first bit of the rCTL valuation function). 
Due to the non-standard semantics of implication in robust logics, this does only work for CTL formulas without implications.  
This is of course not a restriction, as in classical semantics, implication can be derived from disjunction and negation. 

\begin{lemma}\label{lem:CTL to 1st bit of rCTL}
For any CTL state formula $\Phi$ containing no implication, let $\Phi_r$ be the rCTL state formula obtained by dotting all temporal operators in $\Phi$. Then for any state $s$, it holds that
$\ctlsem(s,\Phi) = V_1(s,\Phi_r).$
Consequently, it holds that
$\satset_{\text{CTL}}(\Phi) = \satset(\Phi_r,1111).$
\end{lemma}

\begin{proof}
Applying the definition of the rCTL semantics, we have the following:

\begin{align*}
    V_1(s,p) & = \begin{cases}
        0 & \text{if }p\not \in L(s);\text{ and} \\
        1 & \text{if }p\in L(s),
    \end{cases} \\
    V_1(s,\neg \Phi) & = \begin{dcases}
        0 & \text{if }V_1(s,\Phi)=1;\text{ and} \\
        1 & \text{otherwise},
    \end{dcases}\\
    V_1(s,\Phi \vee \Psi) & =\max\{V_1(s,\Phi),V_1(s,\Psi)\}, \\
    V_1(s,\Phi \wedge \Psi) & =\min\{V_1(s,\Phi),V_1(s,\Psi)\}, \\
    V_1(s,\exists \varphi) & = \max_{\pi \in \paths(s)} V_1(\pi,\varphi), \\
    V_1(s,\forall \varphi) & = \min_{\pi \in \paths(s)} V_1(\pi,\varphi), \\
    V_1(\pi,\Xdot \Phi) & = V_1(\pi[1], \Phi), \\
   V_1(\pi,\Diamonddot \Phi) & = \max_{j\geq 0} V_1(\pi[j],\Phi) \\
    V_1(\pi,\Boxdot \Phi) & =    \min_{j\geq 0} V_1(\pi[j],\Phi) \\
    V_1(\pi,\Phi\Udot \Psi) & = \begin{multlined}[t]
        \max_{j\geq 0} \min \{V_1(\pi[j],\Psi), \\
        \min_{0\leq i < j} V_1(\pi[i],\Phi)\},
    \end{multlined} \\
    V_1(\pi,\Phi\Wdot \Psi) & = \begin{multlined}[t]
        \min_{j\geq 0} \max \{V_1(\pi[j],\Phi), \\
        \max_{0\leq i \leq j} V_1(\pi[i],\Psi)\}.
    \end{multlined}
\end{align*}
Applying these equalities inductively proves that $V_1$ is indeed equal to the valuation $\ctlsem$.
\end{proof}

Hence, rCTL is at least as expressive as CTL. However, the converse is not true, i.e., there exist rCTL formulas that have no equivalent CTL formula. For example, consider the rCTL formula  $\Phi = \forall\Boxdot p$ with truth value $0111$. For a state $s$, we have $s\in \satset(\Phi,0111)$ if and only if for each $\pi\in \paths(s)$, there exists $j$ such that $p\in L(\pi[i])$ for all $i\geq j$, which is equivalent to each path $\pi\in \paths(s)$ satisfying the LTL formula $\Diamond\Box p$. However, the formula $\Diamond\Box p$ can not be expressed in CTL (see Baier and Katoen~\cite{model_checking_book} for details). Therefore, there is no CTL formula $\Psi$ such that $\satset(\Phi,0111) = \satset_{\text{CTL}}(\Psi)$. In total, we obtain the following result.
\begin{theorem}\label{thm:express rCTL CTL}
rCTL is more expressive than CTL.
\end{theorem}

It is known that the expressiveness of LTL and CTL is incomparable. For example, the CTL formula~$\forall \Diamond\forall\Box p$ has no equivalent LTL formula, and the LTL formulas~$\Diamond(p \wedge \X p)$ has no equivalent CTL formula (see Baier and Katoen~\cite{model_checking_book} for details). The same holds for the expressiveness of LTL and rCTL. We just saw that the first bit of the rCTL semantics captures the CTL semantics (for a formula with no implication). Hence, it follows that for the rCTL formula $\forall \Diamonddot\forall\Boxdot p$ (with value $1111$), there is no equivalent LTL formula. Furthermore, one can see that the five-valued semantics does not help in expressing $\varphi = \Diamond(p \wedge \X p)$. Intuitively, a Kripke structure satisfies the formula $\varphi$ if all paths contain a pair of consecutive states where $p$ holds. 
Similarly to the proof of inexpressibility of $\varphi$ in CTL, it can be shown that this property is inexpressible in rCTL as well, as all path formulas are guarded with an existential or universal operator. One can express ``all paths contain a state such that $p$ holds at that state and at all (or some) of its successor'' in rCTL, which is not the same as the property we want. 
Overall, we obtain the following result.

\begin{theorem}\label{thm:express rCTL LTL}
rCTL and LTL have incomparable expressiveness.
\end{theorem}

In the paper on rLTL~\cite{TabuadaN16}, Tabuada and Neider showed that LTL and rLTL are equally expressive. Hence, a direct corollary of \autoref{thm:express rCTL LTL} is the following.
\begin{corollary}\label{corollary:express rCTL rLTL}
rCTL and rLTL have incomparable expressiveness.
\end{corollary}


\subsection{rCTL Model Checking}
\label{sec: rCTL MC}

The classical CTL model checking problem asks whether the computation tree (the tree induced by all its executions) of a given system, satisfies a given CTL specification.
However, in the context of rCTL, this question is more involved due to rCTL's many-valued semantics.
A natural generalization is whether the computation tree satisfies a given property with at least a given value $b_0\in \mathbb{B}_4$. As usual, we model systems by Kripke structures. 
So, the rCTL model checking problem is: for a given finite Kripke structure $M = (S,I,R,L)$, an rCTL formula $\Phi$ and a truth value $b_0\in \mathbb{B}_4$, does $V(s,\Phi) \geq b_0$ hold for all initial states $s\in I$? 

Our rCTL model checking procedure is shown as pseudocode in Algorithm~\ref{alg:rctl model check}.
It is similar to the standard CTL model checking algorithm in that it recursively computes the satisfaction sets $\satset(\Psi,b)$ for each subformula~$\Psi\in \mathrm{Sub}(\Phi)$ and each truth value $b\in \mathbb{B}_4$.
To check whether the Kripke structure satisfies $\Phi$, it is then enough to check whether all initial states belong to $\satset(\Phi,b_0)$.
Note that $\satset(\Psi,0000) = S$ since every state satisfies any rCTL formula $\Psi$ with truth value $0000$.

\begin{algorithm}[!th]
\caption{The rCTL model checking algorithm.}\label{alg:rctl model check}
\textit{Input}: Finite Kripke structure $M$, rCTL formula $\Phi$, and a truth value $b_0\in \mathbb{B}_4$
\begin{algorithmic}
\ForAll{$\Psi\in \mathrm{Sub}(\Phi)$ in increasing size}
	\State $\satset(\Psi,0000) = S$
	\ForAll{$b = 1111 \text{ to } 0001$}
		\State Compute $\satset(\Psi,b)$ as characterized in Table~\ref{table: char of sat}
	\EndFor	
\EndFor
\State \Return $I\subseteq \satset(\Phi,b_0)$
\end{algorithmic}
\end{algorithm}

The key idea of Algorithm~\ref{alg:rctl model check} is to recursively compute the satisfaction sets using a dynamic programming technique. 
More precisely, the satisfaction sets are computed by induction over the structure of $\Phi$ as characterized in Table~\ref{table: char of sat}. 
This characterization is explained in the next paragraphs and proven correct in Lemma~\ref{lemma: char of sat}.
Since $\satset(\Psi,0000) = S$ for any rCTL formula $\Psi$, Table~\ref{table: char of sat} only shows the cases for $b>0000$.

\begin{table*}[t!]
\renewcommand{\arraystretch}{1.5}
\caption{Characterization of the satisfaction sets for rCTL formulas.}
\label{table: char of sat}
\centering
\begin{tabular}{ c l } 
\hline Symbol\mbox{} & $\satset(\cdot,\cdot)$ for rCTL formulas $\Phi$, $\Psi$ and value $b\in \mathbb{B}_4 \setminus \{0000\}$\\
\hline $p\in \mathcal{P}$ & $\satset(p,b) = \{s\in S \mid p\in L(s)\}$ \\
\hline $\vee$ & $\satset(\Phi \vee \Psi,b) = \satset(\Phi,b) \cup \satset(\Psi,b)$\\
\hline $\wedge$ & $\satset(\Phi \wedge \Psi,b) = \satset(\Phi,b) \cap \satset(\Psi,b)$\\
\hline $\neg$ &  $\satset(\neg \Phi, b) = S \setminus \satset(\Phi,1111)$\\
\hline \multirow{2}{*}{$\Rightarrow$} & $\satset(\Phi \Rightarrow \Psi,1111) = \bigcap_b \satset(\Psi,b) \cup (S\setminus \satset(\Phi,b))$ \\ 
 & $\satset(\Phi\Rightarrow \Psi,b) = \satset(\Phi\Rightarrow \Psi,1111) \cup \satset(\Psi,b)$ for any $b\leq 0111$\\
\hline \multirow{2}{*}{$\Xdot$} & $\satset(\exists\Xdot\Phi,b) = \{s\in S \mid \post(s)\cap \satset(\Phi,b)\not = \emptyset\}$\\
& $\satset(\forall\Xdot\Phi,b) = \{s\in S \mid \post(s)\subseteq \satset(\Phi,b)\}$\\
\hline \multirow{2}{*}{$\Diamonddot$} & $\satset(\exists \Diamonddot \Phi,b) = \lfp{T}{F^\exists\big(T,\satset(\Phi,b),S\big)}$\\
& $\satset(\forall \Diamonddot \Phi,b)= \lfp{T}{F^\forall\big(T,\satset(\Phi,b),S\big)}$\\
\hline \multirow{9}{*}{$\Boxdot$} & $\satset(\exists\Boxdot\Phi,1111)= \gfp{T}{F^\exists\big(T,\emptyset,\satset(\Phi,1111)\big)}$\\
& $\satset(\exists\Boxdot\Phi,0111)= \lfp{T_1}{ \gfp{T_2}{ G^\exists(T_1,T_2,\emptyset,\satset(\Phi,0111))}}$\\
& $\satset(\exists\Boxdot\Phi,0011) =\gfp{T_2}{ \lfp{T_1}{ G^\exists(T_1,T_2,\emptyset,\satset(\Phi,0011))}}$ \\
& $\satset(\exists\Boxdot\Phi,0001)= \lfp{T}{F^\exists\big(T,\satset(\Phi,0001),S\big)}$\\
& \vspace*{-0.8em} \\ 
& $\satset(\forall\Boxdot\Phi,1111)= \gfp{T}{F^\forall\big(T,\emptyset,\satset(\Phi,1111)\big)}$\\
& $\satset(\forall\Boxdot\Phi,0111)= \lfp{T_1}{ \gfp{T_2}{ G^\forall(T_1,T_2,\emptyset,\satset(\Phi,0111))}}$\\
& $\satset(\forall\Boxdot\Phi,0011) =\gfp{T_2}{ \lfp{T_1}{ G^\forall(T_1,T_2,\emptyset,\satset(\Phi,0011))}}$\\
& $\satset(\forall\Boxdot\Phi,0001)= \lfp{T}{F^\forall\big(T,\satset(\Phi,0001),S\big)}$\\
\hline \multirow{2}{*}{$\Udot$} & $\satset(\exists (\Phi \Udot \Psi),b)= \lfp{T}{F^\exists\big(T,\satset(\Psi,b),\satset(\Phi,b)\big)}$\\
& $\satset(\forall (\Phi \Udot \Psi),b)= \lfp{T}{F^\forall\big(T,\satset(\Psi,b),\satset(\Phi,b)\big)}$\\
\hline \multirow{9}{*}{$\Wdot$} &  $\satset(\exists(\Phi\Wdot\Psi),1111)= \gfp{T}{F^\exists\big(T,\satset(\Psi,1111),\satset(\Phi,1111)\big)}$\\
& $\satset(\exists(\Phi\Wdot\Psi),0111)= \lfp{T_1}{ \gfp{T_2}{ G^\exists(T_1,T_2,\satset(\Psi,0111),\satset(\Phi,0111))}}$\\ 
& $\satset(\exists(\Phi\Wdot\Psi),0011) =\gfp{T_2}{ \lfp{T_1}{ G^\exists(T_1,T_2,\satset(\Psi,0011),\satset(\Phi,0011))}}$\\
& $\satset(\exists(\Phi\Wdot\Psi),0001)= \lfp{T}{F^\exists\big(T,\satset(\Psi,0001)\cup \satset(\Phi,0001),S\big)}$\\
& \vspace*{-0.8em} \\ 
& $\satset(\forall(\Phi\Wdot\Psi),1111)= \gfp{T}{F^\forall\big(T,\satset(\Psi,1111),\satset(\Phi,1111)\big)}$\\
& $\satset(\forall(\Phi\Wdot\Psi),0111)= \lfp{T_1}{ \gfp{T_2}{ G^\forall(T_1,T_2,\satset(\Psi,0111),\satset(\Phi,0111))}}$\\
& $\satset(\forall(\Phi\Wdot\Psi),0011)=\gfp{T_2}{ \lfp{T_1}{ G^\forall(T_1,T_2,\satset(\Psi,0011),\satset(\Phi,0011))}}$\\
& $\satset(\forall(\Phi\Wdot\Psi),0001)= \lfp{T}{F^\forall\big(T,\satset(\Psi,0001)\cup \satset(\Phi,0001),S\big)}$\\
\hline
\end{tabular}
\end{table*}

To simplify the following presentation of the characterization, we split the discussion into three categories: atomic propositions, Boolean connectives, and temporal operators.

\subparagraph*{Atomic Propositions.} 
The valuation for atomic propositions is defined classically, as in the case of CTL. 
Hence, the satisfaction set $\satset(p,b)$ of an atomic proposition $p\in \mathcal{P}$ with a value $b>0000$ is the set of all states whose label contains~$p$.

\subparagraph*{Boolean Connectives.}
The computation of the satisfaction sets for the Boolean connectives closely follows the semantic definition based on the da~Costa algebra.
Conjunction and disjunction are implemented using the usual intersection and union of sets, respectively. 
The set $\satset(\neg\Phi,b)$ is the complement of all states on which $\Phi$ evaluates to $1111$ (recall that we assume $b > 0000$).
Finally, the implementation of the implication is more involved. 
By definition, the set $\satset(\Phi\Rightarrow\Psi,1111)$ is the set of states $s$ for which $V(s,\Phi)$ is less than $V(s,\Psi)$; in set notation, this is expressed by the intersection of the sets $\satset(\Psi,b) \cup (S\setminus \satset(\Phi,b))$ for each $b\in \mathbb{B}_4$.
For any other truth value $b \leq 0111$, $\satset(\Phi\Rightarrow\Psi,b)$ consists of all states where the implication evaluates to $1111$ or $\Psi$ evaluates to at least $b$.

\subparagraph*{Temporal Operators.}
Now let us explain the characterization of the satisfaction sets for formulas with temporal operators. As the formulas can start with an existential or a universal operator, we discuss the satisfaction sets for them individually.

A state $s$ satisfies the formula $\exists\Xdot\Phi$ with a value of at least $b$ if one of its successors satisfies $\Phi$ with a value of at least $b$. 
Hence, the set $\satset(\exists\Xdot\Phi,b)$ is the set of states $s$ such that one of its successors is in $\satset(\Phi,b)$.
Dually, the set $\satset(\forall\Xdot\Phi,b)$ is the set of states $s$ such that all of its successors are in $\satset(\Phi,b)$.

As for CTL, we use fixed point equations over sets of states to compute satisfaction sets for rCTL formulas with the remaining temporal operators. So, let us first briefly describe some notation and useful properties of fixed point equations over sets of states. A function $F$ that maps a set of states to another set of states is monotonic if $T_1\subseteq T_2$ implies $F(T_1) \subseteq F(T_2)$ for all sets $T_1, T_2$  of states. All monotonic functions have unique least and greatest fixed points~\cite{KnasterTarskiFixedPoint}. Hence, given a monotonic function $F$ (with variable $T$), we write $\lfp{T}{F(T)}$ and $\gfp{T}{F(T)}$ to denote the least fixed point and the greatest fixed point of $F$, respectively. All functions we consider in the following are monotonic.

We begin with formulas of the form~$\exists\Diamonddot\Phi$.
By definition, a state $s$ satisfies $\exists\Diamonddot\Phi$ with a value of at least $b$ if there exists a path from $s$ containing a state that satisfies $\Phi$ with a value of at least $b$. 
Since we are now dealing with paths, we can apply the expansion laws of rLTL~\cite{TabuadaN16}.
In this particular case, we obtain the following statement: a state $s$ satisfies $\exists\Diamonddot\Phi$ with a value of at least $b$ if and only if $s$ satisfies $\Phi$ with a value of at least $b$ or one of its immediate successors satisfies $\exists\Diamonddot\Phi$ with a value of at least $b$. 
Hence, as in CTL, $\satset(\exists\Diamonddot\Phi,b)$ is the smallest subset~$T$ of $S$ satisfying 
$\satset(\Phi,b)\cup \{s\in S\mid \post(s)\cap T\not = \emptyset\}\subseteq T$.
That capture this via fixed point operators, we define the function
\[F^\exists(T,S_1,S_2) = S_1\cup \{s\in S_2 \mid \post(s)\cap T\not = \emptyset\}.\]
So, $F^\exists(T,S_1,S_2)$ contains all states in $S_1$ as well as all states in $S_2$ that have a successor in $T$. Note that this definition is more general than what we need it here, which will be useful for  other temporal operators. But by fixing $S_1 = \satset(\Phi, b)$ and $S_2 = S$, we capture the expansion law of $\exists \Diamonddot $.
Thus, consider the map
\[ T \mapsto F^\exists(T, \satset(\Phi,b), S) \]
mapping sets of states to sets of states. We will prove that its fixed point~$\lfp{T}{F^\exists(T, \satset(\Phi,b), S)}$ is indeed the satisfaction set of $\exists\Diamonddot \Phi$.

Dually, a state $s$ satisfies the formula $\forall\Diamonddot\Phi$ with a value of at least $b$ if every path starting from $s$ contains a state satisfying $\Phi$ with value at least $b$. 
Using analogous arguments, one can show that the set $\satset(\forall\Diamonddot\Phi,b)$ is the least fixed point $\lfp{T}{ F^\forall(T,\satset(\Phi,b),S)}$, where $F^\forall$ is defined as 
\[F^\forall(T,S_1,S_2) = S_1\cup \{s\in S_2 \mid \post(s)\subseteq T\}.\]

Next, we consider formulas of the form~$\exists\Boxdot\Phi$.
The characterization of the set $\satset(\exists\Boxdot\Phi,b)$ is more complex, and we discuss each truth value separately.
Firstly, a state $s$ satisfies $\exists\Boxdot\Phi$ with value $1111$ if there exists a path from $s$ on which every state satisfies $\Phi$ with value $1111$.
By again applying an expansion law similar to that of CTL, this statement is equivalent to $s$ satisfying $\Phi$ with value $1111$ and one of its successors satisfying $\exists\Boxdot\Phi$ with value $1111$. Hence, the set $\satset(\exists\Boxdot\Phi,1111)$ equals the greatest fixed point~$\gfp{T}{F^\exists(T,\emptyset,\satset(\Phi,1111))}$.

Next, a state $s$ satisfies $\exists\Boxdot\Phi$ with a value of at least $0111$ if there exists a path from $s$ on which eventually every state satisfies $\Phi$ with a value of at least $0111$. 
A set of states with such a property can be expressed using nested fixed points as usual (see  Arnold and Niwinski~\cite{Arnold2001RudimentsO} for details).
We will prove that the set $\satset(\exists\Boxdot\Phi,0111)$ is equal to the nested fixed point \[ \lfp{T_1}{ \gfp{T_2}{ G^\exists(T_1,T_2,\emptyset,\satset(\Phi,0111))}},\] where $G^\exists$ is defined as
\begin{align*}
    G^\exists(T_1,T_2,S_1,S_2) = {} &
        S_1 \cup {} \\
        & \{s \in S \mid \post(s)\cap T_1\not = \emptyset\} \cup {} \\
        & \{s\in S_2 \mid \post(s)\cap T_2\not = \emptyset\}.
\end{align*}
Intuitively, the inner greatest fixed point in this nested fixed point represents the property of a path that all states on that path satisfy $\Phi$ with a value of at least $0111$ (similar to the case of $\exists\Boxdot\Phi$ and truth value~$1111$ just discussed).
Then, the outer least fixed point ensures that there exists a path that has a suffix with that property (similar to the case of $\exists\Diamonddot\Phi$ discussed above).

Similarly, a state $s$ satisfies $\exists\Boxdot\Phi$ with a value of at least $0011$ if there exists a path from $s$ on which there exist infinitely many states satisfying $\Phi$ with a value of at least $0011$. Note that the property that a path contains infinitely many states satisfying $\Phi$ (with a value $b$) is the dual of the property that a path contains finitely many states satisfying $\Phi$ (with a value $b$). Hence, similar to the last case, it holds that
\[
    \satset(\exists\Boxdot\Phi,0011) = \begin{multlined}[t] \gfp{T_2}{ \lfp{T_1}{}} \\ G^\exists(T_1,T_2,\emptyset, \satset(\Phi,0011)). \end{multlined}
\]

Finally, a state $s$ satisfies $\exists\Boxdot\Phi$ with a value of at least $0001$ if there exists a path from $s$ containing a state that satisfies $\Phi$ with a value of at least $0001$, which is equivalent to satisfying $\exists\Diamonddot\Phi$ with a value of at least $0001$. Hence, $\satset(\exists\Boxdot\Phi,0001)$ is the least fixed point $\lfp{T}{F^\exists(T,\satset(\Phi,0001),S)}$, as in the case of $\exists\Diamonddot\Phi$.

Analogously, one can characterize $\forall\Boxdot\Phi$ using the fixed points of the functions $F^\forall$ and $G^\forall$, where
\begin{align*}
    G^\forall(T_1,T_2,S_1,S_2) = {} & S_1 \cup {} \\
    & \{s \in S \mid \post(s)\subseteq T_1\} \cup {} \\
    & \{s\in S_2 \mid \post(s)\subseteq T_2\}.
\end{align*}

As the semantics of $\Udot$ mimics the classical semantics, its characterization is generalized from that of $\Diamonddot$, as for CTL.
Hence, its characterization can be obtained using the functions $F^\exists$ and $F^\forall$. 
We describe the case~$\exists \Phi\Udot\Psi$, and the case~$\forall \Phi\Udot\Psi$ is again similar.
A state $s$ satisfies $\exists\Phi\Udot\Psi$ with a value of at least $b$ if there exists a path from $s$ containing a state that satisfies $\Psi$ with a value of at least $b$ and every state before that in the path satisfies~$\Phi$ with a value of at least $b$. 
By applying the expansion law of rLTL~\cite{TabuadaN16}, this statement is equivalent to $s$ satisfying $\Psi$ with a value of at least $b$ or it satisfying~$\Phi$ with a value of at least $b$ and one of its successors satisfying $\exists\Phi\Udot\Psi$ with a value of at least~$b$.
Hence, as in CTL, $\satset(\exists\Phi\Udot\Psi,b)$ is the smallest subset~$T$ of $S$ satisfying 
$\satset(\Psi,b)\cup \{s\in \satset(\Phi,b)\mid \post(s)\cap T\not = \emptyset\}\subseteq T$.
This is captured by the map
\[ T \mapsto F^\exists(T, \satset(\Psi,b), \satset(\Phi,b)). \]
Therefore, the least fixed point
\[
    \lfp{T}{F^\exists(T, \satset(\Psi,b), \satset(\Phi,b))}
\]
of the map is the satisfaction set of $\exists\Phi\Diamonddot \Psi$.

Finally, the semantics of $\Phi\Wdot\Psi$ is also defined using the counting interpretation described in Section~\ref{sec:rCTL-motivation}, similarly to the semantics of $\Boxdot \Phi$. 
However, note that the satisfaction sets for $\Boxdot\Phi$ are characterized only using the satisfaction sets for $\Phi$, whereas the satisfaction sets for $\Phi\Wdot\Psi$ must be characterized using the satisfaction sets of both formulas $\Phi$ and $\Psi$. 
Hence, the characterization for $\Wdot$ can be obtained using the similar fixed points as for $\Boxdot$ but using the satisfaction sets of both formulas $\Phi$ and $\Psi$.

\begin{example}\label{example_mc}
Before proving that the characterization in Table~\ref{table: char of sat} is correct, let us illustrate it on a simple Kripke structure, depicted in Figure~\ref{fig:sat-set-example}.
Continuing the example presented in Section~\ref{sec:Intro}, the Kripke structure demonstrates the interaction between two agents, a robot and office workers (note that for simplicity we consider all workers as one agent).
It captures which agent is present in the dock of the robot (for the sake of readability, we only consider one other location).
Initially, only the robot is present in its dock, captured by the initial state $s_0$ of the Kripke structure.
The robot can continue waiting in its dock, captured by the self-loop in~$s_0$, or it can start performing its task, and, to do so, leave the dock. This is captured by the transition from state~$s_0$ to $s_1$, where there are no agents present at the dock. 
Now, when no agents are present in the dock, represented by state~$s_1$, the office workers can visit the dock, leading to state~$s_2$.
The robot can only return to the dock if it is vacant (encoded by state~$s_1$), i.e., there is no edge from $s_2$ to $s_0$.
Thus, office workers can prevent the robot from returning to its dock, but not continuously, as there is no self-loop in $s_2$: the office worker leaves the dock immediately.

For this Kripke structure, we now compute, for each state, the maximal truth values with which the state 
satisfies the subformulas of $\Phi =\forall\Boxdot \neg H \Rightarrow \forall\Boxdot \exists\Xdot R$.
If this value is $b$ for some state~$s$ and some subformula~$\Psi$, then we have $s \in \satset(\Psi, b')$ for all $b' \le b$ and $s \notin \satset(\Psi, b')$ for all $b' > b$.

These truth values are indicated below the corresponding state in Figure~\ref{fig:sat-set-example}.
\begin{figure*}[th]
\centering
\begin{tikzpicture}[auto]
\node[align=center] (subformulas)  at (-2.5,-2.3) 
{\scriptsize $\begin{aligned}
H&\colon \\ R&\colon\\
\neg H&\colon \\ \exists \Xdot R&\colon\\
\forall\Boxdot\neg H&\colon \\ \forall\Boxdot\exists \Xdot R&\colon\\
\forall\Boxdot\neg H\Rightarrow\forall\Boxdot\exists \Xdot R&\colon
\end{aligned}$
};

\node (0) [state, label={[xshift=0mm,yshift=1.1mm]:$\{R\}$}] at (0,.5) {$s_0$};
\node[align=center] (0-sat)  at (0,-2.3) 
{\scriptsize $\begin{aligned}
&0000\\ &1111\\
&1111\\ &1111\\
&0011\\ &0011\\
&1111
\end{aligned}$
};
\node (1) [state, label={[xshift=0mm,yshift=1.1mm]:$\{\}$}] at (2.5,.5) {$s_1$};
\node[align=center] (1-sat)  at (2.5,-2.3)
{\scriptsize $\begin{aligned}
&0000 \\ &0000\\
&1111 \\ &1111\\
&0011 \\ &0011
\\ &1111
\end{aligned}$
};
\node (2) [state, label={[xshift=0mm,yshift=1.1mm]:$\{H\}$}] at (5,.5) {$s_2$};
\node[align=center] (2-sat)  at (5,-2.3) 
{\scriptsize $\begin{aligned}
&1111 \\ &0000\\ 
&0000 \\ &0000\\ 
&0011 \\ &0011\\
&1111
\end{aligned}$
};
\path [-stealth, thick]
    (-1,.5) edge (0.west)
    (0) edge [bend right] (1)
    (0) edge [loop below] (0)
    (1) edge [bend right] (0)
    (1) edge [loop below] (1)
    (1) edge [bend right] (2)
    (2) edge [bend right] (1);
\end{tikzpicture}
\caption{A Kripke structure that tracks a possible interaction between a robot and office workers. 
Within each state, we mention its identifier.
Above each state, we mention its label. 
Below each state, we mention the maximal valuation that holds in the state for each subformula of $\Phi=\forall\Boxdot \neg H \Rightarrow \forall\Boxdot \exists\Xdot R$.}
\label{fig:sat-set-example}
\end{figure*}
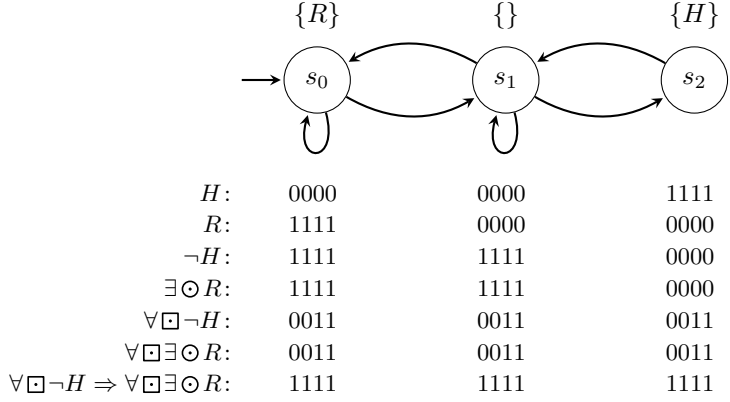

In Figure~\ref{fig:sat-set-example}, observe that $R$ holds with a value of $1111$ in the state $s_0$ and $H$ holds with a value of $1111$ in the state $s_2$ since $s_0$'s label contains $R$ and $s_2$'s label contains $H$.
This is consistent with our characterization of  satisfiable set for atomic propositions, presented in Table~\ref{table: char of sat}: $\satset(R, 1111) =\{s_0\}$ and $\satset(H, 1111)=\{s_2\}$.
Next, observe that the formula $\neg H$ holds in states $s_0$ and $s_1$ with a value of $1111$ since $H$ does not hold in $s_0$ and $s_1$ with a value of $1111$.
This can be seen in the our characterization of negation: $\satset(\neg H, 1111) = S\setminus\satset(H,1111) = S\setminus\{s_2\}=\{s_0,s_1\}$.

The formula $\exists \Xdot R$ holds in the states $s_0$ and $s_1$ with a value of $1111$.
This is because, both $\post(s_0)$ and $\post(s_1)$ contain a state in which $R$ holds with a value of $1111$, namely, the state $s_0$ for both cases.
This is also reflected in our characterization of the next operator: $\satset(\exists \Xdot R, 1111)=\{s\in S \mid \post(s)\cup\satset(R,1111)\neq\emptyset\} = \{s\in S \mid \post(s)\cup\{s_0\}\neq\emptyset\}=\{s_0,s_1\}$.
Also, $\exists \Xdot R$ holds in the state~$s_2$ only with a value of $0000$, as $\post(s_2)$ does not contain a state where $R$ holds with a value larger than $0000$.

Observe that the formula~$\forall\Boxdot\neg H$ holds in all states with a value of $0011$.
This is because every path through the Kripke structure visits infinitely many states where $\neg H$ holds, as $s_2$ does not have a self-loop. Hence, every state satisfies $\forall\Boxdot\neg H$ with at least~$0011$.
This is exactly captured in our characterization for the always operator: $\satset(\forall\Boxdot\neg H,0011) =\gfp{T_2}{ \lfp{T_1}{ G^\forall(T_1,T_2,\emptyset,\satset(\Boxdot\neg H,0011))}}= \gfp{T_2}{ \lfp{T_1}{ G^\forall(T_1,T_2,\emptyset,\{s_0,s_1\})}}=\{s_0,s_1,s_2\}$ (the computation of nested fixed points can be found in Arnold and Niwinski~\cite{Arnold2001RudimentsO}).
On the other hand, from every state there is a path that visits $s_2$ infinitely often, i.e., $H$ holds infinitely often. 
Therefore, no state can satisfy $\forall\Boxdot\neg H$ with $0111$. 
Again, this is captured in our characterization for the always operator: $\satset(\forall\Boxdot\neg H,0111) =\lfp{T_1}{ \gfp{T_2}{ G^\forall(T_1,T_2,\emptyset,\satset(\Boxdot\neg H,0111))}}= \lfp{T_1}{ \gfp{T_2}{ G^\forall(T_1,T_2,\emptyset,\{s_0,s_1\})}}=\emptyset$.
Thus, $\forall\Boxdot\neg H$ holds in all states with a maximal value of $0011$.

In a fashion similar to $\forall\Boxdot\neg H$, the formula $\forall\Boxdot\exists\Xdot R$ holds with a value of $0011$ in all the states.
This is also reflected in our characterization for the always operator.

Now, since the maximal value of $\forall\Boxdot\neg H$ is equal to that of $\forall\Boxdot\exists\Xdot R$ in all the states, $\Phi=\forall\Boxdot \neg H \Rightarrow \forall\Boxdot \exists\Xdot R$ holds in all the states with a value of $1111$.
Also, our characterization of implication states $\satset(\forall\Boxdot \neg H \Rightarrow \forall\Boxdot \exists\Xdot R, 1111)= \bigcap_b(\satset(\forall\Boxdot \exists\Xdot R,b)\cup S\setminus\satset(\forall\Boxdot \neg H,b))=\{s_0,s_1,s_2\}$. %
\mbox{}\hfill\mbox{}\exampleend
\end{example}

\begin{lemma}\label{lemma: char of sat}
The characterization of the satisfaction sets in Table~\ref{table: char of sat} is correct.
\end{lemma}

\begin{proof}
Let $M = (S,I,R,L)$ be a given Kripke structure and $b\in \mathbb{B}_4\setminus\{0000\}$. Now, we show that every equation in Table~\ref{table: char of sat} using a case-by-case analysis.

\begin{itemize}
\def\labelitemi{\textbullet}
\setlength\itemsep{0.5em}

\item $\begin{aligned}[t]
&s\in \satset(p,b)\\
&\iff V(s,p)\geq b>0000\\
&\iff V(s,p) = 1111 \\
&\iff p\in L(s).
\end{aligned}$

\item$\begin{aligned}[t]
&s\in \satset(\Phi\vee\Psi,b)\\
&\iff V(\Phi\vee\Psi)\geq b\\
&\iff \max\{V(s,\Phi),V(s,\Psi)\}\geq b\\
&\iff V(s,\Phi)\geq b \text{ or } V(s,\Psi)\geq b\\
&\iff s\in \satset(\Phi,b) \text{ or } s\in \satset(\Psi,b)\\
&\iff s\in \satset(\Phi,b) \cup \satset(\Psi,b).
\end{aligned}$

\item$\begin{aligned}[t]
&s\in \satset(\Phi\wedge\Psi,b)\\
&\iff V(\Phi\wedge\Psi)\geq b\\
&\iff \min\{V(s,\Phi),V(s,\Psi)\}\geq b\\
&\iff V(s,\Phi)\geq b \text{ and } V(s,\Psi)\geq b\\
&\iff s\in \satset(\Phi,b) \text{ and } s\in \satset(\Psi,b)\\
&\iff s\in \satset(\Phi,b) \cap \satset(\Psi,b).
\end{aligned}$

\item $\begin{aligned}[t]
&s\in \satset(\neg\Phi,b)\\
&\iff V(s,\neg \Phi) \geq b \geq 0001\\
&\iff \overline{V(s,\Phi)} = 1111\\ 
&\iff V(s,\Phi) \not= 1111\\ 
&\iff s\in S\setminus \satset(\Phi,1111).
\end{aligned}$

\item $\begin{aligned}[t]
&s\in \satset(\Phi\Rightarrow\Psi,1111)\\
&\iff \big(V(s,\Phi) \rightarrow V(s,\Psi)\big) = 1111\\ 
&\iff V(s,\Phi) \leq V(s,\Psi) \\
&\iff \forall b\in \mathbb{B}_4 \colon s \not \in  \satset(\Phi,b)\setminus \satset(\Psi,b) \\ 
&\iff \forall b\in \mathbb{B}_4 \colon s \in  \big(S\setminus \satset(\Phi,b)\big)\cup \satset(\Psi,b)\\
&\iff s \in \bigcap_b \satset(\Psi,b) \cup (S\setminus \satset(\Phi,b)).
\end{aligned}$

Similarly, for some $b\leq 0111,$

$\begin{aligned}[t]
&s\in \satset(\Phi\Rightarrow\Psi,b)\\
&\iff \big(V(s,\Phi) \rightarrow V(s,\Psi)\big) \geq  b\\
&\iff V(s,\Phi) \leq V(s,\Psi) \text{ or } V(s,\Psi) = b\\
&\iff s\in \satset(\Phi\Rightarrow \Psi,1111) \cup \satset(\Psi,b).
\end{aligned}$

\item   $\begin{aligned}[t]
&s\in \satset(\exists\Xdot \Phi,b)\\
&\iff \exists\pi\in \paths(s) \colon V(\pi,\Xdot\Phi)\geq b\\
&\iff \exists\pi\in \paths(s) \colon V(\pi[1],\Phi)\geq b \\
&\iff \exists s'\in \post(s) \colon V(s',\Phi)\geq b \\
&\iff \post(s)\cap \satset(\Phi,b)\not = \emptyset\\
\end{aligned}$

and

$\begin{aligned}[t]
&s\in \satset(\forall\Xdot \Phi,b)\\
&\iff \forall \pi\in \paths(s) \colon V(\pi,\Xdot\Phi)\geq b \\
&\iff \forall \pi\in \paths(s) \colon V(\pi[1],\Phi)\geq b \\
&\iff \forall s'\in \post(s) \colon V(s',\Phi)\geq b \\
&\iff \post(s)\subseteq \satset(\Phi,b).
\end{aligned}$

\end{itemize}
It remains to consider the temporal operators eventually, always, until, and release. 
Here, we need to prove that the fixed-point characterizations presented are correct. 
For most cases, the technical core of the arguments are standard characterizations of safety (a state formula holds at every state of a given path), co-Büchi (a state formula holds almost always on a given path),  Büchi (a state formula holds infinitely often on a given path), and reachability (a state formula holds in at least one state of a given path) conditions. We present several cases in detail and refer for the other ones to the book by Arnold and Niwinski~\cite{Arnold2001RudimentsO} for more details.

\begin{itemize}
\item
$\begin{aligned}[t]
&s\in \satset(\exists(\Phi\Udot\Psi),b)\\
& \iff \exists\pi\in \paths(s) \colon V(\pi,\Phi\Udot\Psi)\geq b\\
& \iff \begin{multlined}[t]
    \exists\pi\in \paths(s), \exists j\geq 0, \forall i< j \colon \\
    V(\pi[j],\Psi)\geq b \wedge V(\pi[i],\Phi) \geq b
\end{multlined} \\
& \iff \begin{multlined}[t]
    \big(V(s,\Psi)\geq b\big) \vee \big(V(s,\Phi)\geq b \wedge {} \\
    \exists s'\in \post(s) \colon V(s',\exists(\Phi\Udot\Psi))\geq b\big)
\end{multlined} \\
& \iff \begin{multlined}[t]
    s\in \satset(\Psi,b) \cup \{s'\in \satset(\Phi,b) \mid {} \\
    \post(s')\cap \satset(\exists(\Phi\Udot\Psi,b))\not = \emptyset\}.
\end{multlined}
\end{aligned}$

Hence, $\satset(\exists(\Phi\Udot\Psi),b)$ is a fixed point of the function $T \mapsto F^\exists(T,\satset(\Psi,b),\satset(\Phi,b))$. 
Now, we only need to show that it is indeed the least fixed point. Suppose $T'$is another fixed point of that function. If $s_0\in \satset(\exists(\Phi\Udot\Psi),b)$, then there exists a path $\pi = s_0s_1s_2\cdots$ and some $j>0$ such that $V(s_j,\Psi)\geq b$ and $V(s_i,\Phi) \geq b$ for all $0\leq i<j$. Then:
\begin{itemize}
\setlength\itemsep{0em}
\def\labelitemii{\textbf{-}}
\item $s_j\in \satset(\Psi,b) \subseteq T'$;
\item $s_{j-1}\in T'$, since $s_j\in \post(s_{j-1}) \cap T'$ and $s_{j-1}\in \satset(\Phi,b)$;
\item $s_{j-2}\in T'$, since $s_{j-1}\in \post(s_{j-2}) \cap T'$ and $s_{j-2}\in \satset(\Phi,b)$;
\item Applying this argument repeatedly yields $s_0\in T'$.
\end{itemize}
So, $\satset(\exists(\Phi\Udot\Psi),b) \subseteq T'$. Therefore, $\satset(\exists(\Phi\Udot\Psi),b)$ is the least fixed point of $ T \mapsto F^\exists(T,\satset(\Psi,b),\satset(\Phi,b))$. Similarly, it can be shown that $\satset(\forall(\Phi\Udot\Psi),b)$ is the least fixed point of $T \mapsto F^\forall(T,\satset(\Psi,b),\satset(\Phi,b))$.

\item In the following, we use $\mathtt{true}$ as syntactic sugar for some tautology, e.g., $p \vee \neg p$. Then, $\Diamonddot \Phi $ is equivalent to $ \mathtt{true}\Udot \Phi$ and we have $\satset(\mathtt{true},b) = S$. Hence,

$\begin{aligned}
\satset(\exists\Diamonddot \Phi,b) & = \satset(\exists (\mathtt{true}\Udot\Phi))\\ 
&= \lfp{T}{\begin{multlined}[t] F^\exists\big(T,\satset(\Phi,b), \\ \satset(\mathtt{true},b)\big) \end{multlined}}\\
&= \lfp{T}{F^\exists\big(T,\satset(\Phi,b),S\big)}.\\
\end{aligned}$

Similarly, we have
\[ \satset(\forall\Diamonddot\Phi,b) = \lfp{T}{F^\forall\big(T,\satset(\Phi,b),S\big)}, \]
as claimed.

\item $\begin{aligned}[t]	
&s\in \satset(\exists\Boxdot \Phi,1111) \\
&\iff \exists\pi\in \paths(s)\colon V(\pi,\Boxdot\Phi)=1111\\
&\iff \exists\pi\in \paths(s), \forall i\geq 0 \colon V(\pi[i],\Phi) = 1111\\
&\iff \begin{multlined}[t]
    s\in \satset(\Phi,1111) \wedge {} \\
    \big(\post(s) \cap \satset(\exists\Boxdot\Phi,1111) \not =\emptyset\big)
\end{multlined} \\
& \iff \begin{multlined}[t]
    s\in \satset(\Phi,1111) \cap {} \\
    \{s'\mid \post(s') \cap \satset(\exists\Boxdot\Phi,1111) \not =\emptyset\}.
\end{multlined}
\end{aligned}$

Hence, $\satset(\exists\Boxdot \Phi,1111)$ is a fixed point of the function $T \mapsto F^\exists(T,\emptyset,\satset(\Phi,1111))$.
Now, we only need to show that it is indeed the greatest fixed point. Now, suppose $T'$ is another fixed point. If $s_0\in T'$, then

\begin{itemize}
\setlength\itemsep{0em}
\item since $s_0\in T'$, there exists a state $s_1\in \post(s_0)\cap T'$;
\item since $s_1\in T'$, there exists a state $s_2\in \post(s_0)\cap T'$;
\end{itemize}

Applying this argument iteratively yields that there exists a path $s_0s_1\cdots$ starting from $s$ such that $V(s_i,\Phi) = 1111$ for each $i\geq 0$. Hence, $s_0\in \satset(\exists\Boxdot \Phi,1111)$, which implies $T'\subseteq \satset(\exists\Boxdot \Phi,1111)$. Therefore, $\satset(\exists\Boxdot \Phi,1111)$ is the greatest fixed point of $T \mapsto F^\exists(T,\emptyset,\satset(\Phi,1111))$.

Similarly, the following holds

$\begin{aligned}[t]	
&s\in \satset(\exists\Boxdot\Phi,0111)\\ 
&\iff \exists\pi\in \paths(s), V(\pi,\Boxdot\Phi)\geq 0111\\
&\iff \begin{multlined}[t]
    \exists \pi\in \paths(s), \exists j\geq 0, \forall i\geq j \colon \\
    V_2(\pi[i],\Phi) = 1
\end{multlined}\\
&\iff \text{$\pi$ visits $\satset(\Phi,0111)$ eventually always.}\\
&\text{Moreover, }s\in \satset(\exists\Boxdot\Phi,0011)\\ 
&\iff \exists\pi\in \paths(s), V(\pi,\Boxdot\Phi)\geq 0011\\
&\iff \begin{multlined}[t]
    \exists \pi\in \paths(s), \forall j\geq 0, \exists i>j \colon \\
    V_3(\pi[i],\Phi) = 1
\end{multlined}\\
&\iff \text{$\pi$ visits $\satset(\Phi,0011)$ infinitely often.}\\
\end{aligned}$

A path visiting a set eventually always or infinitely often can be written in terms of nested fixed points as claimed (see Arnold and Niwinski~\cite{Arnold2001RudimentsO} for details).

$\begin{aligned}[t]
&\text{Finally, }s\in \satset(\exists\Boxdot\Phi,0001)\\ 
&\iff \exists\pi\in \paths(s), V(\pi,\Boxdot\Phi)\geq 0001.\\
&\iff \exists\pi\in \paths(s), \exists i\geq 0 \colon V_4(\pi[i],\Phi) = 1\\
&\iff s\in \satset(\exists\Diamonddot\Phi,0001).
\end{aligned}$

Hence,
\[ \satset(\exists\Boxdot\Phi,0001) = \lfp{T}{F^\exists\big(T,\satset(\Phi,0001),S\big)}, \]
as claimed.

Analogously, one can show the claimed results for $\forall\Boxdot\Phi$. 

\item For the operator $\Wdot$, the claim can be shown using arguments similar to those for $\Boxdot$.\qedhere
\end{itemize}
\end{proof}

Algorithm~\ref{alg:rctl model check} computes $5\cdot \lvert \mathrm{sub}(\Phi)\rvert$ satisfaction sets following the subformula ordering. 
Using the standard fixed point iterations~\cite{fixed-point-iteration}, the nested fixed points of depth two can be computed in time~$\mathcal{O}(NK)$ on a Kripke structure with $N$ vertices and $K$ transitions~\cite{fixed-point-buchi}. So, we obtain the following.

\begin{theorem}
Given an rCTL formula $\Phi$ and a Kripke structure with $N$ states and $K$ transitions, the rCTL model checking problem can be solved in time~$\mathcal{O}(NK\lvert \Phi\rvert)$.
\end{theorem}

Note that the CTL model checking algorithm also takes polynomial time in the size of the formula and the number of transitions of the Kripke structure~\cite{CTL-model-checking}. Hence, both model checking problems are in $\mathrm{PTIME}$.
Moreover, a lower bound  of the rCTL model checking problem can be derived from the $\mathrm{PTIME}$ lower bound of CTL model checking~\cite{DBLP:conf/aiml/Schnoebelen02} and Lemma~\ref{lem:CTL to 1st bit of rCTL}.
In total, we obtain the following result, showing that the CTL and rCTL model checking problems have the same asymptotic complexity.

\begin{corollary}
The model checking problem for rCTL is $\mathrm{PTIME}$-complete.
\end{corollary}

\subsection{rCTL and the Modal \texorpdfstring{\boldmath$\mu$}{mu}-calculus}
\label{sec: mucalculus}

In the previous section, we have seen that one can solve the rCTL model checking problem by computing least and greatest fixed points. 
In this section, we show that every rCTL formula can be translated into an equivalent formula of the modal $\mu$-calculus~\cite{mu-calculus}, i.e., modal logic with least and greatest fixed points. 
This is not necessarily surprising, as most temporal logics can be translated into the modal $\mu$-calculus~\cite{logic2mucalculus}.
However, the result is very useful to settle the complexity of satisfiability and synthesis, which we achieve by reductions to satisfiability and synthesis for the modal $\mu$-calculus. 

We begin by reviewing the basic definitions of the modal $\mu$-calculus. It consists of state formulas only, which are constructed from atomic propositions with Boolean connectives, the temporal operators $\exists\X$ and $\forall\X$, as well as the least~($\mu$) and the greatest~($\nu$) fixed point operator.

Formally, given a set $\mathcal{P}$ of atomic propositions and a set $\mathcal{PV}$ of atomic proposition variables, $\mu$-calculus formulas are given by the grammar
\begin{multline*}
\Phi \Coloneqq p \mid y \mid \Phi \vee \Phi \mid \Phi \wedge \Phi \mid \neg \Phi \mid \Phi \Rightarrow \Phi \mid \\
\exists\X \varphi \mid \forall\X \varphi \mid \mu y. \Phi \mid \nu y. \Phi,
\end{multline*}
where $p\in \mathcal{P}$ and $y \in \mathcal{PV}$. 
As usual, we require that in subformulas of the form~$\mu y. \Phi $ and $\nu y.\Phi$, every free occurrence of $y$ in $\Phi$ is under the scope of an even number of negations.
For further details, we refer the reader to standard literature on this topic, e.g., Grädel, Thomas, and Wilke~\cite{Automata_Book}.

Unlike temporal logics, the semantics of the $\mu$-calculus is naturally defined using satisfaction sets. 
Given a Kripke structure $M=(S,I,R,L)$, the satisfaction sets are defined with respect to a variable function $\variablefunc\colon \mathcal{PV}\rightarrow 2^S$ that maps each atomic proposition variable to a set of states.
Moreover, for a subset $T\subseteq S$, let $\variablefunc[y\rightarrow T]$ denote the variable function that maps $y$ to $T$ while preserving the value of $\variablefunc$ for every other input.

Given a variable function $\variablefunc$ and a $\mu$-calculus formula $\Phi$, let $\musatset(\Phi)$ denote the set of states satisfying $\Phi$ with respect to $\variablefunc$. These sets are defined recursively using the least and greatest fixed points, as shown in Table~\ref{table: mu-calulus sat}.
The functions the fixed point operators are applied to are monotonic since every occurrence of a fixed point variable is under an even number of negations.
Hence, the fixed points all exist. 

For $\mu$-calculus sentences, i.e., formulas without free variables, the satisfaction sets~$\satset_\mu^\variablefunc$ are independent of $\variablefunc$.
Hence, we drop the parameter~$\variablefunc$ from the notation whenever possible.

\begin{table*}[t!]
\centering
\caption{Characterization of the satisfaction sets for $\mu$-calculus formulas.}\label{table: mu-calulus sat}
\renewcommand{\arraystretch}{1.5}
\begin{tabular}{ c l } 
\hline Symbol\mbox{} & $\musatset(\cdot)$ for $\mu$-calculus formulas $\Phi$, $\Psi$\\
\hline $p\in \mathcal{P}$ & $\musatset(p) = \{s\in S \mid p\in L(s)\}$ \\
\hline $\vee$ & $\musatset(\Phi \vee \Psi) = \musatset(\Phi) \cup \musatset(\Psi)$\\
\hline $\wedge$ & $\musatset(\Phi \wedge \Psi) = \musatset(\Phi) \cap \musatset(\Psi)$\\
\hline $\neg$ &  $\musatset(\neg \Phi) = S \setminus \musatset(\Phi)$\\
\hline $\Rightarrow$ & $\musatset(\Phi \Rightarrow \Psi) = \musatset(\neg\Phi) \cup \musatset(\Psi)$\\
\hline $\exists\X$ & $\musatset(\exists\X\Phi) = \{s\in S \mid \post(s)\cap \musatset(\Phi)\not = \emptyset\}$\\
\hline $\forall\X$ & $\musatset(\forall\X\Phi) = \{s\in S \mid \post(s)\subseteq \musatset(\Phi)\}$\\
\hline $y\in \mathcal{PV}$ & $\musatset(y) = \variablefunc(y)$ \\
\hline $\mu y$ & $\musatset(\mu y. \Phi) = \lfp{T}{\satset_\mu^{\variablefunc[y\rightarrow T]}(\Phi)}$\\
\hline $\nu y$ & $\musatset(\nu y. \Phi) = \gfp{T}{\satset_\mu^{\variablefunc[y\rightarrow T]}(\Phi)}$\\
\hline
\end{tabular}
\end{table*}

We now show that, like other temporal logics, rCTL can also be translated into the modal $\mu$-calculus. As before, we say an rCTL formula $\Phi$ with a truth value $b\in \mathbb{B}_4$ is equivalent to a $\mu$-calculus sentence~$\Phi'$ if for every Kripke structure it holds that $\satset(\Phi,b) = \satset_\mu(\Phi')$. Then we have the following result.
\begin{theorem}\label{thm:mucalctransl}
For every rCTL formula and truth value, there is an equivalent $\mu$-calculus sentence of linear size.
\end{theorem}

\begin{proof}
We show that there exists a mapping $t$ that assigns to every rCTL formula $\Phi$ and truth value $b\in \mathbb{B}_4$ an equivalent $\mu$-calculus formula $t(\Phi,b)$. We define this mapping recursively, starting with the atomic rCTL formulas. 
In the following proof, we use $\mathtt{true}$ as syntactic sugar for an arbitrary tautology of the $\mu$-calculus, e.g., $p \vee \neg p$.

First of all, for any rCTL formula $\Phi$ with truth value $0000$, a trivial equivalent $\mu$-calculus formula is $t(\Phi,0000) = \mathtt{true}$.
Furthermore, comparing the characterization of the satisfaction sets of rCTL and the $\mu$-calculus (Tables~\ref{table: char of sat} and \ref{table: mu-calulus sat}), one can see that for a Boolean combination of rCTL formulas $\Phi$ and $\Psi$ with any truth value $b\in \mathbb{B}_4\setminus\{0000\}$, the following recursive translations indeed results in an equivalent $\mu$-calculus formula:
\begin{align*}
    t(p, b) & = p \text{ for each }p\in\mathcal{P},\\
    t(\Phi \vee \Psi, b) & = t(\Phi, b) \vee t(\Psi, b), \\
    t(\Phi \wedge \Psi, b) & = t(\Phi, b) \wedge t(\Psi, b), \\
    t(\neg \Phi, b) & = \neg t(\Phi,1111). \\
    \intertext{Moreover, we have}
    t(\Phi\Rightarrow\Psi,1111) & = \bigwedge_{b} t(\Psi,b)\vee \neg t(\Phi,b),\\
    \intertext{and}
    t(\Phi\Rightarrow\Psi,b) & = t(\Phi\Rightarrow\Psi,1111) \vee t(\Psi,b)
\end{align*}
for any $b\leq 0111$.

The rCTL formulas with the next operator are captured by applying the $\mu$-calculus operators $\exists\X$ and $\forall\X$ as follows for $b\in \mathbb{B}_4\setminus\{0000\}$:
\begin{align*}
    t(\exists\Xdot\Phi,b) & = \exists\X t(\Phi,b),\\
    t(\forall\Xdot\Phi,b) & = \forall\X t(\Phi,b).
\end{align*}

For rCTL formulas with other temporal operators, the satisfaction sets (in Table~\ref{table: char of sat}) are defined using fixed points of functions $F^\exists$, $F^\forall$, $G^\exists$, and $G^\forall$. Hence, we first give $\mu$-calculus formulas that capture these functions. Note that if the sets $T,S_1,S_2$ are the satisfaction sets of the rCTL formulas $\Phi_t,\Phi_1,\Phi_2$ with truth values $b_t,b_1,b_2$, respectively, then it holds that
\begin{align*}
    F^\exists(T,S_1,S_2) & = S_1 \cup \{s\in S_2 \mid \post(s)\cap T \neq \emptyset\} \\
    & = \begin{multlined}[t]
        S_1 \cup (S_2 \cap \{s\in S\mid \post(s)\cap T \neq \emptyset\})
    \end{multlined}\\
    & = \begin{multlined}[t]
        \satset(\Phi_1,b_1) \cup {} \\
        \big(\satset(\Phi_2,b_2) \cap \satset(\exists\Xdot\Phi_t,b_t)\big).
    \end{multlined}
\end{align*}
Now, suppose that the $\mu$-calculus formula $t(\Phi_1,b_1)$ is equivalent to the rCTL formula $\Phi_1$ with truth value~$b_1$, and the $\mu$-calculus formula $t(\Phi_2,b_2)$ is equivalent to the rCTL formula $\Phi_2$ with truth value~$b_2$.
Then, we have 
\begin{multline*}
  F^\exists \big(T,\satset(\Phi_1,b_1),\satset(\Phi_2,b_2)\big) = {} \\
  \satset_\mu^{\variablefunc[y\rightarrow T]} \big(t(\Phi_1,b_1)\vee (t(\Phi_2,b_2)\wedge \exists\X y)\big).  
\end{multline*}
Therefore, for $\mu$-calculus formulas $\Phi_1'$ and $\Phi_2'$, the function $F^\exists$ can be represented by the following $\mu$-calculus formula containing $y$ as a free variable:
\[F^\exists_\mu(y,\Phi_1',\Phi_2') = \Phi_1' \vee (\Phi_2'\wedge \exists\X y).\]
Hence, using Table~\ref{table: char of sat}, one can see that an equivalent $\mu$-calculus formula for the rCTL formula $\exists\Diamonddot\Phi$ with truth value $b\in\mathbb{B}_4\setminus \{0000\}$ is the following:
\[t(\exists\Diamonddot \Phi, b) = \mu y. F^\exists_\mu(y,t(\Phi,b),\mathtt{true}).\]

Similarly, the functions $F^\forall$, $G^\exists$, and $G^\forall$ can be represented by the following $\mu$-calculus formulas:
\begin{align*}
    F^\forall_\mu(y,\Phi_1',\Phi_2') & = \Phi_1' \vee (\Phi_2'\wedge \forall\X y),\\
    G^\exists_\mu(y_1,y_2,\Phi_1',\Phi_2') & = \exists\X y_1 \vee \Phi_1' \vee (\Phi_2' \wedge \exists\X y_2),\\
    G^\forall_\mu(y_1,y_2,\Phi_1',\Phi_2') & = \forall\X y_1 \vee \Phi_1' \vee (\Phi_2' \wedge \forall\X y_2).
\end{align*}
Now, for an rCTL formula with temporal operators $\Diamonddot$, $\Boxdot$, $\Udot$, and $\Wdot$, we obtain an equivalent $\mu$-calculus formula of linear size from the characterization of satisfaction sets given in Table~\ref{table: char of sat} by replacing the functions and the satisfaction sets of subformulas with corresponding $\mu$-calculus formulas.
\end{proof}

\new{While it is true that every CTL formula can be transformed into an equivalent alternation-free (with alternation depth 1, as defined in \cite{Bradfield2018}) $\mu$-calculus formula, it's important to note that the constructed $\mu$-calculus formulas for rCTL formulas typically have an alternation depth of at most 2. This limitation arises from the presence of two-depth alternation for some rCTL operators, such as $\exists\Boxdot$ with value $0011$ as illustrated in Table~\ref{table: char of sat}.
Furthermore, as the model checking problem for $\mu$-calculus formulas with alternation depth $d$ can be solved in time $\mathcal{O}(n^{d+1})$~\cite{EmersonJ91,Bradfield2018}, one can also solve rCTL model checking in cubic time by reducing it to $\mu$-calculus model checking.
}

Let us conclude by mentioning that the converse of Theorem~\ref{thm:mucalctransl} does not hold: rCTL is strictly less expressive than the modal $\mu$-calculus.
This follows from a stronger result presented to be presented in Section~\ref{subsec:rctlstarexpressiveness}.

\subsection{rCTL Satisfiability}
\label{sec:rctlsat}

This section considers the satisfiability problem for rCTL, which is: given an rCTL formula $\Phi$ and a truth value $b_0\in\mathbb{B}_4$, does there exist a Kripke structure $M = (S,I,R,L)$ such that $I\subseteq \satset(\Phi,b_0)$?
The next theorem settles the complexity of the rCTL satisfiability problem.

\begin{theorem}
\label{thm:rctlsat}
The satisfiability problem for rCTL is $\mathrm{EXPTIME}$-complete.
\end{theorem}

\begin{proof}
The upper bound is obtained by translating a given rCTL formula and a given truth value into an equivalent $\mu$-calculus formula of linear size (see Theorem~\ref{thm:mucalctransl}) and then checking the resulting formula for satisfiability.
Since the satisfiability problem for the $\mu$-calculus (defined as expected) is $\mathrm{EXPTIME}$-complete~\cite{mu-calculus-CTL*-sat}, rCTL satisfiability is in $\mathrm{EXPTIME}$ as well.

The matching lower bound already holds for CTL satisfiability (again defined as expected)~\cite{CTL-sat}, which, due to Lemma~\ref{lem:CTL to 1st bit of rCTL}, reduces to rCTL satisfiability.
\end{proof}

Moreover, since every satisfiable formula of the $\mu$-calculus has a model of exponential size~\cite{mu-calculus-model}, the same is true for rCTL.

\begin{corollary}
\label{cor:rCTL-satisfiability-size-model}
Every satisfiable rCTL-formula has a model of exponential size.
\end{corollary}

There are satisfiable CTL formulas that have only models of at least exponential size~\cite{quirky}.\footnote{Note that the exponential lower bound is shown with respect to the \emph{length} of the formula, i.e., the number of nodes of the syntax tree of the formula. In contrast, we measure the size of a formula by the number of distinct subformulas, i.e., the number of distinct subtrees of the syntax tree. Thus, our complexity measure might be smaller, which only strengthens the lower bound.} Thus, the upper bound in Corollary~\ref{cor:rCTL-satisfiability-size-model} is tight.

Also, note that the asymptotic complexity of the rCTL satisfiability
problem and the size of a model matches that of CTL.

\subsection{rCTL Synthesis}
\label{sec:rctlsynthesis}

We now turn to the problem of rCTL synthesis.
The synthesis problem asks, given an rCTL specification on the input-output behavior of a system, whether there is a system satisfying the specification, and, if yes, compute one.
As a preparatory step, let us first introduce the required notation.

Let $\mathcal{P} = \mathcal{I} \cup \mathcal{O}$ be the disjoint union of a set $\mathcal I$ of input propositions and a set $\mathcal O$ of output propositions. 
A strategy is a mapping~$\strat\colon (2^\mathcal{I})^* \rightarrow 2^\mathcal{O}$. 
Note that finite automata with input alphabet~$2^{\mathcal{I}}$ and output alphabet~$2^{\mathcal{O}}$ (with Mealy or Moore semantics) can be used to implement strategies. 
We call such strategies finite-state and measure their size in the number of states of the automaton.

A strategy~$\strat$ induces an infinite Kripke structure~$M_\strat = (S, I, R, L)$  with $S = (2^\mathcal{I})^+$, $I = 2^\mathcal{I}$, $R = \{(w, wa) \mid w \in (2^\mathcal{I})^+, a \in 2^\mathcal{I}\}$, and $L(wa) = a \cup \strat(wa)$.

We say that $\strat$ realizes an rCTL formula~$\Phi$ with at least value~$b_0 \in \mathbb{B}_4$ if $V(s,\Phi) \ge b_0$ for all initial states~$s$ of $M_\strat$.
Further, a rCTL formula is realizable with at least value~$b_0$ if there is a strategy that realizes it with at least~$b_0$.
The rCTL synthesis problem is: given an rCTL formula~$\Phi$ and a truth value~$b_0 \in \mathbb{B}_4$, is $\Phi$ realizable with at least~$b_0$?
The next theorem settles its complexity.

\begin{theorem} \label{thm:rCTL-synthesis}
The rCTL synthesis problem is $\mathrm{EXPTIME}$-complete.
\end{theorem}

\begin{proof}
The upper bound is obtained by translating a given rCTL formula and a given truth value into an equivalent $\mu$-calculus formula of linear size (see Theorem~\ref{thm:mucalctransl}) and then checking the resulting formula for realizability (which is defined as expected).
Since the synthesis problem for the $\mu$-calculus is $\mathrm{EXPTIME}$-complete~\cite{mu-calculus-synthesis}, rCTL synthesis is in $\mathrm{EXPTIME}$ as well.

The matching lower bound already holds for CTL synthesis~\cite{branchingtimesynthesis}, which, due to Lemma~\ref{lem:CTL to 1st bit of rCTL}, reduces to rCTL synthesis.
\end{proof}

As every realizable formula of the $\mu$-calculus is realized by a finite-state strategy of exponential size, which can be computed in exponential time~\cite{branchingtimesynthesis}, the same is true for rCTL.

\begin{corollary} \label{cor:rCTL-synthesis-size-strategy}
If an rCTL-formula~$\varphi$ is realizable with at least~$b_0$, then one can compute, in exponential time, an exponentially-sized finite-state strategy realizing $\varphi$ with at least~$b_0$.
\end{corollary}

There are realizable CTL formulas that are only realized by finite-state strategies of exponential size: this follows from the exponential lower bound on the size of model (see the discussion below Corollary~\ref{cor:rCTL-satisfiability-size-model}) and the fact that satisfiability can be reduced to synthesis~\cite{branchingtimesynthesis}. Hence, the upper bound in Corollary~\ref{cor:rCTL-synthesis-size-strategy} is tight.

Finally, note that Theorem~\ref{thm:rCTL-synthesis} and Corollary~\ref{cor:rCTL-synthesis-size-strategy} imply that the rCTL synthesis problem again has the same asymptotic complexity as the one for CTL.

\new{
\section{Review of CTL*}
\label{sec:ctl-star}

In this section, we briefly review the syntax and semantics of CTL*, which we then robustify to obtain robust CTL*.

\subsection{Syntax}

Unlike CTL, CTL* allows path quantifiers $\exists$ and $\forall$ to be arbitrarily nested with temporal operators. The syntax of CTL* state formulas is the same as in CTL. Moreover, CTL* path formulas are similar to LTL formulas. Consequently, CTL* state formulas over $\mathcal{P}$ are formed according to the grammar
\[\Phi \Coloneqq p \mid \Phi \vee \Phi \mid \Phi \wedge \Phi \mid \neg \Phi \mid \Phi \Rightarrow \Phi \mid \exists \varphi \mid \forall \varphi,\]
where $p\in \mathcal{P}$ and $\varphi$ is a path formula.
CTL* path formulas are formed according to the grammar
\begin{multline*}
    \normalfont \varphi \Coloneqq \Phi \mid \varphi \vee \psi\mid \varphi \wedge \psi \mid \neg \varphi\mid \varphi \Rightarrow \psi \mid \\
    \X \varphi\mid \Diamond\varphi\mid \Box\varphi\mid \varphi \U \psi \mid \varphi \W \psi.
\end{multline*}

\subsection{Semantics}

Let $M$ be a Kripke structure, $\Phi$, $\Psi$ two CTL* state formulas, and $\varphi$, $\psi$ two CTL* path formulas.
For a state $s$, the CTL* semantics $\ctlstarsem(s,\Phi)$ is defined as the CTL semantics (see Subsection~\ref{sec:CTL-semantics}).
For a path $\pi$, the semantics is analogous to the LTL semantics via a valuation function~$\ctlstarsem$:
\begin{align*}
    \ctlstarsem(\pi,\Phi) & = \ctlstarsem(\pi[0],\Phi), \\
    \ctlstarsem(\pi,\varphi \vee \psi) & = \max\{\ctlstarsem(\pi,\varphi),\ctlstarsem(\pi,\psi)\}, \\
    \ctlstarsem(\pi,\varphi \wedge \psi) & = \min\{\ctlstarsem(\pi,\varphi),\ctlstarsem(\pi,\psi)\}.\\
    \ctlstarsem(\pi,\neg \varphi) & = 1 - \ctlstarsem(\pi, \varphi).\\
	     \ctlstarsem(\pi, \varphi \Rightarrow \psi)& = \begin{cases}
         1 &\hspace{-.2cm} \text{if }\ctlstarsem(\pi, \varphi)\le\\
         &\hspace{-.2cm} \ctlstarsem(\pi, \psi); \text{and} \\
         \ctlstarsem(\pi, \psi)&\hspace{-.2cm}\text{otherwise, }\\
     \end{cases}\\
 	\ctlstarsem(\pi,\X \varphi) & = \ctlstarsem(\pi[1..], \varphi).\\
    \ctlstarsem(\pi,\Diamond \varphi) & = \max_{i\geq 0} \ctlstarsem(\pi[i],\varphi).\\
    \ctlstarsem(\pi,\Box \varphi) & = \min_{i\geq 0} \ctlstarsem(\pi[i], \varphi),\\
    \ctlstarsem(\pi,\varphi\U \psi) & = \begin{multlined}[t]
        \max_{j\geq 0} \min \{\ctlstarsem(\pi[j..],\psi), \\
        \min_{0\leq i < j} \ctlstarsem(\pi[i..],\varphi)\},
    \end{multlined}\\
    \ctlstarsem(\pi,\varphi\W \psi) & =  \begin{multlined}[t]
            \min_{j\geq 0}\max \{\ctlstarsem(\pi[j..],\varphi), \\
            \max_{0\leq i \leq j} \ctlstarsem(\pi[i..],\psi)\},
        \end{multlined} \\
\end{align*}

}

\section{Robust CTL*}
\label{sec:rctl-star}

In this section, we present the robust version of CTL*, named robust CTL*, which combines the features of rCTL and rLTL.
We show that rCTL* is more expressive than both.
In addition, we present an rCTL* model checking algorithm and address the rCTL* satisfiability and synthesis problems.

\subsection{Syntax}

Like CTL*, robust CTL* allows path quantifiers $\exists$ and $\forall$ to be arbitrarily nested with temporal operators. The syntax of rCTL* state formulas is the same as in rCTL and CTL*. Moreover, rCTL* path formulas are similar to rLTL formulas, with the only difference being the use of arbitrary rCTL* state formulas as atoms. Consequently, rCTL* state formulas over $\mathcal{P}$ are formed according to the grammar
\[\Phi \Coloneqq p \mid \Phi \vee \Phi \mid \Phi \wedge \Phi \mid \neg \Phi \mid \Phi \Rightarrow \Phi \mid \exists \varphi \mid \forall \varphi,\]
where $p\in \mathcal{P}$ and $\varphi$ is a path formula.
rCTL* path formulas are formed according to the grammar
\begin{multline*}
    \normalfont \varphi \Coloneqq \Phi \mid \varphi \vee \psi\mid \varphi \wedge \psi \mid \neg \varphi\mid \varphi \Rightarrow \psi \mid \\
    \Xdot \varphi\mid \Diamonddot\varphi\mid \Boxdot\varphi\mid \varphi \Udot \psi \mid \varphi \Wdot \psi.
\end{multline*}

Again, the set of subformulas of a state formula~$\Phi$ is denoted by $\mathrm{Sub}(\Phi)$ and the size of a formula is defined as the number of its syntactically different subformulas.

\subsection{Semantics}

As in CTL*, the semantics for rCTL* state and path formulas are analogous to the rCTL and rLTL semantics, respectively.
In what follows, let $M$ be a Kripke structure, $\Phi$, $\Psi$ two rCTL* state formulas, and $\varphi$, $\psi$ two rCTL* path formulas.

For a state $s$, the rCTL* semantics $V(s,\Phi)$ is then the same as the rCTL semantics (see Section~\ref{sec:rCTL-semantics}).

For a path $\pi$, the semantics is analogous to the rLTL semantics (cf.\ Tabuada and Neider~\cite{TabuadaN16}) via a valuation function~$V$ (note that, for notational convenience, we use the letter $V$ both the rCTL and the rCTL* valuation function):
\begin{align*}
    V(\pi,\Phi) & = V(\pi[0],\Phi), \\
    V(\pi,\varphi \vee \psi) & = \max\{V(\pi,\varphi),V(\pi,\psi)\}, \\
    V(\pi,\varphi \wedge \psi) & = \min\{V(\pi,\varphi),V(\pi,\psi)\}.\\
    V(\pi,\neg \varphi) & = \overline{V(\pi, \varphi)}.\\
	V(\pi,\varphi \Rightarrow \psi) & = V(\pi,\varphi)\rightarrow V(\pi,\psi).\\ 
 	V(\pi,\Xdot \varphi) & = V(\pi[1..], \varphi).\\
    V(\pi,\Diamonddot \varphi) & = \max_{i\geq 0} V(\pi[i],\varphi).\\
    V(\pi,\Boxdot \varphi) & = (b_1,b_2,b_3,b_4) \text{ where} \\
        b_1 & = \min_{i\geq 0} V_1(\pi[i], \varphi),\\
        b_2 & = \max_{j\geq 0} \min_{i\geq j} V_2(\pi[i], \varphi),\\
        b_3 & = \min_{j\geq 0} \max_{i\geq j} V_3(\pi[i], \varphi),\\
        b_4 & = \max_{i\geq 0} V_4(\pi[i], \varphi)), \\
    V(\pi,\varphi\Udot \psi) & = \begin{multlined}[t]
        \max_{j\geq 0} \min \{V(\pi[j..],\psi), \\
        \min_{0\leq i < j} V(\pi[i..],\varphi)\},
    \end{multlined}\\
    V(\pi,\varphi\Wdot \psi) & = (b_1,b_2,b_3,b_4) \text{ where} \\
        b_1 & = \begin{multlined}[t]
            \min_{j\geq 0}\max \{V_1(\pi[j..],\varphi), \\
            \max_{0\leq i \leq j} V_1(\pi[i..],\psi)\},
        \end{multlined} \\
        b_2 & = \begin{multlined}[t]
            \max_{k\geq 0} \min_{j\geq k}\max \{V_2(\pi[j..],\varphi), \\
            \max_{0\leq i \leq j} V_2(\pi[i..],\psi)\},
        \end{multlined} \\
        b_3 & = \begin{multlined}[t]
            \min_{k\geq 0} \max_{j\geq k}\max \{V_3(\pi[j..],\varphi), \\
            \max_{0\leq i \leq j} V_3(\pi[i..],\psi)\},
        \end{multlined} \\
        b_4 & = \begin{multlined}[t]
            \max_{j\geq 0}\max \{V_4(\pi[j..],\varphi), \\
            \max_{0\leq i \leq j} V_4(\pi[i..],\psi)\}.
        \end{multlined}
\end{align*}

Before studying the properties of rCTL*, let us illustrate the difference between rCTL and rCTL* using an example.

\begin{example}
\label{example_rctlstar}
Continuing our running example from Section\ref{sec:Intro}, we illustrate how the rCTL* formula $ \forall(\Boxdot \neg H\Rightarrow \Boxdot \exists\Xdot R)$ is different from the rCTL formula $ \forall\Boxdot \neg H\Rightarrow \forall\Boxdot \exists\Xdot R$ from Examples~\ref{example_rctl} and \ref{example_mc}. 
Recall that $\neg H$ states that office workers are not at the robot's dock and $\exists\Xdot R$ states that the robot can return to its dock in one time step. Assume $\forall(\Boxdot \neg H\Rightarrow \Boxdot \exists\Xdot R)$ evaluates to $1111$. Then the formula $\Boxdot \neg H\Rightarrow \Boxdot \exists\Xdot R$ must evaluate to $1111$ for each path. Hence, the following holds:
\begin{itemize}
\item If $\neg H$ holds at every state in a path $\pi$, then $V(\pi,\Boxdot \neg H)$ evaluates to $1111$. Hence, by the rCTL* semantics, $V(\pi,\Boxdot \exists\Xdot R)$ must also evaluate to $1111$. That means, $\exists\Xdot R$ also holds at every state in $\pi$. Hence, in any path, if office workers never visit the dock, then from every state, the robot can return to its dock in one time step.

\item Similarly, if $\neg H$ holds eventually always for some path $\pi$, then $V(\pi,\Boxdot \neg H)$ evaluates to $0111$. Then, by the rCTL* semantics, $V(\pi,\Boxdot \exists\Xdot R)$ evaluates to $0111$ or higher. Hence, $\exists\Xdot R$ also needs to hold eventually always in $\pi$. Therefore, if office workers visit the dock a few times and never visit it again in a path, then from any state in that path, the robot can return to its dock eventually.

\item Similarly, if $\neg H$ holds at infinitely (finitely) many states in some path $\pi$, then $\exists\Xdot R$ needs to hold at infinitely (finitely) many states in $\pi$. 
\end{itemize}
As we can see, the rCTL* formula~$\forall(\Boxdot \neg H\Rightarrow \Boxdot \exists\Xdot R)$ captures the robustness property for every path separately, whereas the rCTL formula~$\forall\Boxdot \neg H\Rightarrow \forall\Boxdot \exists\Xdot R$ captures the robustness property jointly for all paths starting from a state.

\begin{figure}[b!]
	\centering
	\begin{tikzpicture}
		\node[player0, label={below:$\{ \neg H, \exists\Xdot R \}$}] (0) at (0, 0) {$s_0$};
		\node[player0, label={below:$\{ \neg H \}$}] (1) at (-2, 0) {$s_1$};
		\node[player0, label={below:$\{ \}$}] (2) at (2, 0) {$s_2$};
		
		\path[initial] (0.north) -- +(0, 0.5);
		\path[->] (0) edge (1) edge (2);
		\path[->] (1) edge[loop above] ();
		\path[->] (2) edge[loop above] ();		
	\end{tikzpicture}
	\caption{The Kripke structure for Example~\ref{example_rctlstar}. States are labeled with the formulas that hold at that state with truth value~$1111$.}
	\label{fig:difference}
\end{figure}
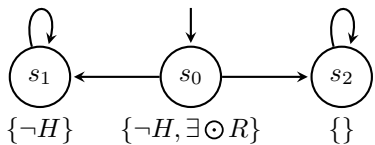

To understand the difference, let us consider the Kripke structure $M$ with initial state $s_0$ as shown in Figure~\ref{fig:difference} (where transitions are depicted by edges). Suppose the set of states that satisfy (with value~$1111$) the state formulas~$\neg H$ and $\exists\Xdot R$ are $\{s_0,s_1\}$ and $\{s_0,s_2\}$, respectively (as shown by the labels in the figure). 

There are only two paths starting from $s_0$, i.e., $\pi_1 = s_0 s_1 s_1 \cdots$ and $\pi_2 = s_0 s_2 s_2 \cdots$. Since $\neg H$ holds at every state in the path~$\pi_1$, we have $V(\pi_1,\Boxdot \neg H) = 1111$. Moreover, since $\neg H$ holds only at the first state in the path~$\pi_2$, we have $V(\pi_2,\Boxdot\neg H) = 0001$. Hence, $V(s_0,\forall\Boxdot\neg H) = \min_{i\in\{1,2\}} V(\pi_i,\Boxdot\neg H) = 0001$. Similarly, since $\exists\Xdot R$ holds only at the first state of each path, we have $V(\pi_1,\Boxdot\exists\Xdot R) = V(\pi_2,\Boxdot\exists\Xdot R) = 0001$. Hence, $V(s_0,\forall\Boxdot\exists\Xdot R) = 0001$. Therefore, it holds that $V(s,\forall\Boxdot\neg H \Rightarrow \forall\Boxdot\exists\Xdot R) = 1111$ according to the rCTL semantics.

Now, let us consider the rCTL* formula~$\forall(\Boxdot\neg H\Rightarrow \Boxdot\exists\Xdot R)$.
As we have $V(\pi_1,\Boxdot\exists\Xdot R) = 0001 < V(\pi_1,\Boxdot\neg H)$, it holds that $V(\pi_1, \Boxdot\neg H\Rightarrow \Boxdot\exists\Xdot R) = 0001$. Similarly, we have $V(\pi_2,\Boxdot\neg H\Rightarrow\Boxdot\exists\Xdot R) = 1111$. Hence, we have $V(s,\forall(\Boxdot\neg H\Rightarrow \Boxdot\exists\Xdot R)) = 0001$ according to the rCTL* semantics.

Recall that the rCTL formula~$\forall\Boxdot\neg H\Rightarrow \forall\Boxdot\exists\Xdot R$ evaluates at state~$s$ to $1111 \neq 0001$.
This is the case because both of the paths do not satisfy $\Boxdot\neg H\Rightarrow \Boxdot\exists\Xdot R$ with value $1111$ individually, but collectively, the state $s_0$ satisfies $\forall\Boxdot\neg H\Rightarrow \forall\Boxdot\exists\Xdot R$. 
\mbox{}\hfill\mbox{}\exampleend
\end{example}

\subsection{Expressiveness of rCTL*}
\label{subsec:rctlstarexpressiveness}

The satisfaction sets and the equivalence between two formulas in rCTL* are defined as for rCTL.
As we can see, rCTL* is an extension of both rCTL and rLTL. Therefore, it subsumes both rCTL and rLTL (and hence, it also subsumes \new{CTL and} LTL). 
Furthermore, by Corollary~\ref{corollary:express rCTL rLTL}, there exist rCTL formulas that are not expressible in rLTL and vice versa.
In total, we obtain the following result.

\begin{theorem}
rCTL* is more expressive than rLTL, rCTL, \new{CTL,} and LTL. 
\end{theorem}

Using the same idea as in Lemma~\ref{lem:CTL to 1st bit of rCTL}, one can recover the CTL* semantics of a formula with no implication from the first component of the rCTL* semantics. Conversely, using the same arguments as for the analogous result for rLTL~\cite[Proposition~$5$]{TabuadaN16}, one can translate each rCTL* formula into four CTL* formulas that captures the four components of the rCTL* semantics.
Hence, we obtain the following result. 

\begin{theorem}\label{thm:CTLstarandrCTLstar}
CTL* and rCTL* are equally expressive.
\end{theorem}

\begin{proof}
For any CTL* formula $\Phi$ containing no implication, let $\Phi_r$ be the rCTL* formula obtained by dotting all temporal operators in $\Phi$. Then for any state $s$, it holds that
$\ctlsem(s,\Phi) = V_1(s,\Phi_r)$, which is shown as the analogous result for CTL and rCTL (see the proof of Lemma~\ref{lem:CTL to 1st bit of rCTL}). 
Consequently, it holds that $\satset_{\text{CTL*}}(\Phi) = \satset(\Phi_r,1111).$ Furthermore, as $\Phi \Rightarrow \Psi$ is equivalent to $\Psi\vee\neg\Phi$ in CTL*, every CTL* formula can be rewritten as a formula containing no implication. Therefore, for every CTL* formula, there is an equivalent rCTL* formula with respect to the truth value~$1111$.

For the other direction, we define a mapping $t$ that assigns to every rCTL* state formula $\Phi$ and truth value~$b$ an equivalent CTL* formula $t(\Phi,b)$.
Furthermore, $t$ maps every rCTL* path formula $\varphi$ and a truth value $b$ to $t(\varphi, b)$ such that for every path $\pi$,
\[V_{CTL^*}(\pi,t(\varphi,b)) = 1 \text{ if and only if } V(\pi,\varphi) \geq b. \]
Again, for $b=0000$, we can define $t(\Phi, b) = \mathtt{true}$ for every state formula~$\Phi$, where $\mathtt{true}$ is an arbitrary tautology of CTL*, e.g., $p \vee \neg p$.

In the following, we assume $b > 0000$.
Then, for state formulas $\Phi$ and $\Psi$, the mapping $t$ is defined inductively as follows:
\begin{align*}
    t(p, b) & = p \text{ for any $p\in\mathcal{P}$}, \\
    t(\Phi \vee \Psi, b) & = t(\Phi, b) \vee t(\Psi, b), \\
    t(\Phi \wedge \Psi, b) & = t(\Phi, b) \wedge t(\Psi, b), \\
    t(\neg \Phi, b) & = \neg t(\Phi,1111). \\
    \intertext{Moreover, we define}
    t(\Phi\Rightarrow\Psi,1111) & = \bigwedge_{b > 0000} t(\Psi,b)\vee \neg t(\Phi,b), \\
    \intertext{and}
    t(\Phi\Rightarrow\Psi,b) & = t(\Phi\Rightarrow\Psi,1111) \vee t(\Psi,b)
\end{align*}
for each $b\leq 0111$.

For Boolean combinations of path formulas, the mapping $t$ can be defined analogously as for state formulas. For path formulas $\varphi$ and $\psi$ with temporal operators, $t$ is defined as follows:

\begin{align*}
    t(\exists\varphi,b) & = \exists t(\varphi,b), \\
    t(\forall\varphi,b) & = \forall t(\varphi,b), \\
    t(\Xdot \varphi,b) & = \X t(\varphi,b), \\
    t(\Diamonddot \varphi,b) & = \Diamond t(\psi,b), \\
    t( \Boxdot \varphi,1111) & = \Box t(\varphi,1111), \\
    t(\Boxdot \varphi,0111) & = \Diamond \Box t(\varphi,0111), \\
    t(\Boxdot \varphi,0011) & = \Box \Diamond t(\varphi,0011), \\
    t(\Boxdot \varphi,0001) & = \Diamond t(\varphi,0001), \\
    t(\varphi \Udot \psi,b) & = t(\varphi,b) \U t(\psi,b), \\
    t(\varphi \Wdot \psi,1111) & = t(\varphi,1111)\W t(\psi,1111), \\
    t(\varphi \Wdot \psi,0111) & = \Diamond \Box t(\varphi,0111)\vee \Diamond t(\psi,0111), \\
    t(\varphi \Wdot \psi,0011) & = \Box \Diamond t(\varphi,0011)\vee \Diamond t(\psi,0011), \\
    t(\varphi \Wdot \psi,0001) & = \Diamond t(\varphi,0001)\vee \Diamond t(\psi,0001).
\end{align*}

A structural induction shows that the translation~$t$ has the desired property.
\end{proof}

Although we do not require the result, let us just mention that Theorem~\ref{thm:CTLstarandrCTLstar} also gives a translation of rCTL* into the modal $\mu$-calculus: rCTL* can be translated into CTL*, which can be translated into the modal $\mu$-calculus~\cite{logic2mucalculus}.
Conversely, the modal $\mu$-calculus is strictly more expressive than CTL* (see, e.g., Demri, Goranko, and Lange~\cite[Chapter 10]{temporallogicsbook}). Hence, it is also strictly more expressive than rCTL* (due to Theorem~\ref{thm:CTLstarandrCTLstar}) and rCTL (which is a fragment of rCTL*).

\subsection{rCTL* Model Checking}
\label{sec:rctlstarmc}

The model checking problem for rCTL* is analogous to that of rCTL: for a given finite Kripke structure $M = (S,I,R,L)$, an rCTL* formula $\Phi$ and a truth value $b_0\in \mathbb{B}_4$, does $V(s,\Phi) \geq b_0$ hold for all initial states $s\in I$? 
One way of solving the rCTL* model checking problem is by applying the translation from rCTL* into CTL* (see Theorem~\ref{thm:CTLstarandrCTLstar}) and then apply a CTL* model checking algorithm. 

Here, we will present an alternative approach via a combination of rCTL and rLTL model checking. This approach is analogous to the classical CTL* model checking algorithm, is a combination of CTL and LTL model checking. 
In practice, the choice for one algorithm over the other depends on whether one wants to apply an CTL* model checker or an rLTL model checker.

As in rCTL, for the rCTL* model checking, we use the characterization of the satisfaction sets. $\satset(\Phi,b)$ can be computed using Table~\ref{table: char of sat} for every state formula $\Phi$ which is either an atomic proposition or can be expressed as a Boolean combination (conjunction, negation, etc.) of two subformulas. Otherwise, we use an rLTL model checking algorithm to compute $\satset(\Phi,b)$ for a state formula starting with a path quantifier.

Let us first go through the basic concepts of rLTL and its model checking algorithm. As we have described earlier, rCTL* is an extension of rLTL. Both rCTL* path formulas and rLTL formulas are defined using almost the same syntax, with the only difference being the use of state formulas as atoms in rCTL*. Moreover, the valuation $V$ for rLTL formulas is defined the same way as it is defined for rCTL* path formulas. The rLTL model checking problem is: given a Kripke structure~$M$, an rLTL formula $\varphi$, and a truth value~$b_0\in \mathbb{B}_4$, determine whether for all initial states $s$ and all $\pi \in \paths(s)$, it holds that $V(\pi,\varphi)\ge b_0$.\footnote{Actually, the original definition by Tabuada and Neider is slightly more general~\cite{TabuadaN16}.} To solve the rLTL model checking problem, Tabuada and Neider~\cite{TabuadaN16} have provided an algorithm to compute a generalized B\"uchi automaton (see Gr\"adel, Thomas and Wilke~\cite{Automata_Book} for a definition) recognizing all traces over the alphabet~$2^{\mathcal{P}}$ satisfying a given formula with a value $b\in B$ for a given set $B\subseteq \mathbb{B}_4$, as formalized below.
\begin{lemma}[Tabuada and Neider~\cite{TabuadaN16}]\label{lem:rLTL buchi}
Given an rLTL formula $\varphi$, and a set of truth values $B\subseteq \mathbb{B}_4$, one can construct a generalized B\"uchi automaton $A_{\varphi,B}$ with $\mathcal{O}(5^{\lvert \varphi\rvert})$ states and $\mathcal{O}(\lvert \varphi\rvert)$ accepting sets that recognizes all paths $\pi$ such that $V(\pi,\varphi) \in B$.
\end{lemma}
One can now solve the rLTL model checking problem by first translating $M$ into a B\"uchi automaton~$A_M$ accepting exactly the traces labeling the paths of $M$ starting in an initial state. Then, one determines whether $L(A_M)\cap L(A_{\varphi,\{ b \in \mathbb{B}_4 \mid b < b_0 \}})$ is empty.

Coming back to computing $\satset(\Phi,b)$ for $\Phi$ starting with a path quantifier, let us consider $\Phi = \forall\varphi$. Observe that $s\in \satset(\forall\varphi,b)$ if and only if $V(s,\forall\varphi)\geq b$. Further, $V(s,\forall\varphi)\geq b$ if and only if $V(\pi,\varphi)\geq b \text{ for all }\pi\in \paths(s)$.
The basic idea is now to replace all maximal proper state subformulas $\Psi$ of $\varphi$ by fresh atomic propositions $a_{\Psi}$ and use the rLTL model checking algorithm to compute all the states from which all paths satisfy the rLTL formula $\varphi$ with value at least~$b$. However, we need to make a minor modification in the construction of the B\"uchi automaton of Lemma~\ref{lem:rLTL buchi} such that for each $a_{\Psi}$, it holds that $V(s,a_{\Psi}) \geq b$ whenever $s\in \satset(\Psi,b)$ and $V(s,a_{\Psi}) < b$ whenever $s\not\in \satset(\Psi,b)$. This can be done by initializing these atomic propositions with the required truth value.

Similarly, we compute $\satset(\exists\varphi,b)$ by the rLTL model checking algorithm using the observation that $s\not \in \satset(\exists\varphi,b)$ if and only if $V(\pi,\varphi)< b \text{ for all }\pi\in \paths(s)$.

Now, one can solve the rCTL* model checking problem using Algorithm~\ref{alg:rctl model check}. However, the time complexity of the algorithm is not the same as in rCTL since the computation of $\satset$ uses the rLTL model checking algorithm, which takes exponential time in the size of the formula and the Kripke structure (Tabuada and Neider~\cite{TabuadaN16}). Hence, the time complexity of the rCTL* model checking algorithm is dominated by the time complexity of the rLTL model checking algorithm. 

Altogether, our algorithm runs in polynomial space (as rLTL model checking is in $\mathrm{PSPACE}$~\cite{TabuadaN16}).
A matching lower bound already holds for CTL*~\cite{CTL*-model-checking}.

\begin{theorem}
The rCTL* model checking problem is $\mathrm{PSPACE}$-complete.
\end{theorem}

The CTL* model checking problem is also $\mathrm{PSPACE}$-complete~\cite{CTL*-model-checking}. Hence, both the CTL* and the rCTL* model checking problem have the same asymptotic complexity.

\subsection{rCTL* Satisfiability}
\label{sec:rctlstarsat}

This section considers the satisfiability problem for rCTL*, which is: for a given rCTL* formula~$\Phi$ and truth value $b_0\in\mathbb{B}_4$, does there exist a Kripke structure $M = (S,I,R,L)$ such that $I\subseteq \satset(\Phi,b_0)$? 

\begin{theorem}
The satisfiability problem for rCTL* is $\mathrm{2EXPTIME}$-complete.
\end{theorem}

\begin{proof}
Both the upper and the lower bound follow immediately from Theorem~\ref{thm:CTLstarandrCTLstar} and the fact that CTL* satisfiability (which is defined as expected) is $\mathrm{2EXPTIME}$-complete~\cite{mu-calculus-CTL*-sat}.
The linear translation from rCTL* to CTL* yields the upper bound while the linear translation from CTL* to rCTL* yields the lower bound.
\end{proof}

Furthermore, as every satisfiable CTL* formula has a model of doubly-exponential size
 (see, e.g., Demri, Goranko, and Lange~\cite[Chapter 15]{temporallogicsbook}), the same is true for rCTL*.

\begin{corollary}
\label{cor:rCTLstar-satisfiability-size-model}
Every satisfiable rCTL* formula has a model of doubly-exponential size.
\end{corollary}

There are satisfiable CTL* formulas that have only models of doubly-exponential size (see, e.g., Demri, Goranko, and Lange~\cite[Chapter 15]{temporallogicsbook}). 
Hence, the upper bound in Corollary~\ref{cor:rCTLstar-satisfiability-size-model} is tight.

Also, note again that the asymptotic complexity of the rCTL* satisfiability problem and the size of a model matches that of CTL*.

\subsection{rCTL* Synthesis}
\label{sec:rctlstarsynthesis}

The notions of a strategy realizing an rCTL formula with at last value~$b_0$ and an rCTL formula being realizable can be generalized to rCTL*. 
Then, the rCTL* synthesis problem is defined analogously to the rCTL synthesis problem: given an rCTL* formula~$\Phi$ and a truth value~$b_0 \in \mathbb{B}_4$, is $\Phi$ realizable with value at last~$b_0$?

\begin{theorem}
The rCTL* synthesis problem is $\mathrm{2EXPTIME}$-complete.
\end{theorem}

\begin{proof}
The lower bound again follows immediately from CTL* synthesis (again defined as expected) being $\mathrm{2EXPTIME}$-complete~\cite{branchingtimesynthesis} and the fact that the CTL* semantics is a special case of rCTL* (see Theorem~\ref{thm:CTLstarandrCTLstar}).

Here, the upper bound follows from the converse translation, i.e., given a rCTL* formula~$\Phi$ and a truth value~$b_0 \in \mathbb{B}_4$, Theorem~\ref{thm:CTLstarandrCTLstar} allows us to construct (in linear time) a CTL* formula~$t(\Phi,{b_0})$ that is equivalent to $\Phi$ with respect to $b_0$: a strategy realizes $\Phi$ with at least~$b_0$ if and only if it realizes $t(\Phi,{b_0})$. 
As CTL* synthesis is in $\mathrm{2EXPTIME}$~\cite{branchingtimesynthesis}, we obtain the desired upper bound.
\end{proof}

As every realizable CTL* formula is realized by a finite-state strategy of doubly-exponential size~\cite{branchingtimesynthesis}, which can be computed in doubly-exponential time, the same is true for rCTL*.

\begin{corollary}
\label{cor:rCTLstar-synthesis-size-strategy}
If an rCTL*-formula $\Phi$ is realizable with at least~$b_0$, then one can compute, in doubly-exponential time, a doubly-exponentially-sized finite-state strategy realizing $\varphi$ with at least~$b_0$.
\end{corollary}

There are realizable LTL formulas (and therefore CTL* formulas, as LTL is a fragment of CTL*) that are only realized by finite-state strategies with at least doubly-exponentially many states~\cite{pnuelirosner}. Hence, the doubly-exponential upper bound in Corollary~\ref{cor:rCTLstar-synthesis-size-strategy} is tight.

Perhaps unsurprisingly at this point, the asymptotic complexity of the rCTL* synthesis problem and the tight bound on the size of a finite-state strategy realizing an rCTL* specification match those of CTL*.

\new{
\section{Related Works}\label{sec:related}

Numerous efforts have been made to formalize the concept of robustness in cyber-physical systems within the framework of formal methods. This section offers an extensive yet not exhaustive overview of various formalizations of robustness. We initiate our discussion with a series of approaches that necessitate designers to provide additional information alongside their desired specifications.

In the work by Bloem et al.~\cite{BloemGHJ09}, two quantitative robustness concepts are combined into a unified framework for robust synthesis. The first concept, known as robustness for safety, examines how frequently assumptions and guarantees are violated, with a requirement that their ratio remains bounded by a parameter $k\in\mathbb{N}$ (referred to as $k$-robustness). This counting process relies on error functions supplied by the designer. The second concept, robustness for liveness, deals with specifications in the form of $\bigwedge_{i\in I}\,\Diamond\Box p_i \implies \bigwedge_{j\in J}\,\Diamond\Box q_j$, where $p_i$ and $q_j$ are atomic propositions. It compares the number of violated assumptions to the number of violated guarantees. While our semantics can distinguish between different ways of specification violations, it does not distinguish between the violation of one assumption and multiple assumptions. Consequently, this second approach is not directly comparable to the one proposed here. Furthermore, we do not make a distinction between safety and liveness properties.

In another work by Bloem et al.~\cite{bloem2019synthreacsys}, a distinct framework for robust synthesis is introduced, which does not encompass the previously mentioned framework. This framework considers various notions of robustness, such as a system being robust if it satisfies a guarantee even when a finite number or even all of its inputs are hidden or misread, or when the assumptions are violated either finitely or infinitely often. Many of these notions align with our notion, and our definition of robustness allows systems to satisfy weaker guarantees whenever the assumptions are also weakened, making it more general. However, we cannot directly compare our approach with the notions of robustness in \cite{bloem2019synthreacsys} that involve counting the number of violations since our semantics distinguishes only between zero, finite, and infinite violations of a specification.

In the work of Rodionova et al.~\cite{rodionova2016logicfiltering}, a connection is established between MTL/LTL and Linear Time-Invariant (LTI) filtering. Specifically, it is demonstrated that LTI filtering corresponds to MTL if addition and multiplication are interpreted as max and min operations, and if true and false are interpreted as one and zero, respectively. Different filtering kernels are employed to express weaker or stronger interpretations of the same formula, placing the burden on the designer to choose kernels and use multiple semantics to reason about how weakening assumptions affect guarantees.

In contrast to the approaches mentioned above, which necessitate designers to provide robustness metrics, our approach simplifies the designer's task by requiring only the desired specification without the need for additional metrics. This simplification is especially beneficial as it may not always be clear which quantitative metric leads to the desired qualitative behavior.


In the realm of software systems, Zhang et al.~\cite{zhang2020behavrobsoftsys} define robustness as the largest set of deviating environmental behaviors under which the system still guarantees a desired property. Therefore, robustness is defined as the set of all deviations under which a system continues to satisfy that property. While this work focuses on computing robustness rather than characterizing it, it is possible that certain temporal deviations could be expressed in our semantics. Additional noteworthy works, although not directly comparable to the methods described here, include Chaudhuri et al.~\cite{chaudhuri2010contanalysisprograms} and Majumdar et al.~\cite{majumdar2009symbrobanalysis}, which consider continuity properties of software expressed by the requirement that a deviation in a program's input causes a proportional deviation in its output. However, these notions of robustness only apply to the Turing model of computation and not to the reactive model of computation employed in this paper.

Several works have explored the robustness of specifications when reasoning over real-valued, continuous-time signals, with prominent examples being Fainekos et al.~\cite{fainekos2009robtemplogcontsignals}, Donze et al.~\cite{donze2010robtemplogrealval}, Akazaki et al.~\cite{AkazakiH15}, Abbas et al.~\cite{AbbasPM19}, and Mehdipour et al.~\cite{MehdipourVB19}. In these works, specific choices made when crafting many-valued semantics are not discussed in detail. Notably, the notion of ``time robustness'' in~\cite{donze2010robtemplogrealval} is somewhat similar to that of our semantics in that it measures the time needed for the truth value of a formula to change. However, in this line of work, robustness is derived from the real-valued nature of signals, whereas our semantics reasons over the more classical setting of discrete-time and Boolean-valued signals, with robustness derived from the temporal evolution of these signals. 
Consequently, these works on real-valued signals~\cite{fainekos2009robtemplogcontsignals,donze2010robtemplogrealval,AkazakiH15,AbbasPM19,MehdipourVB19} and their extensions can be considered orthogonal and complementary to our approach.

Another relevant approach involving multi-valued extensions of LTL is presented in Almagor et al.~\cite{almagor2016quality}. This work introduces two quantitative extensions of LTL, one with propositional quality operators denoted as LTL$[\mathcal{F}]$, parameterized by a set $\mathcal{F}$ of functions over $[0,1]$, and another with discounting operators termed LTL$^{disc}[\mathcal{D}]$, parameterized by a set $\mathcal{D}$ of discounting functions. Both logics employ a many-valued variant of LTL to reason about quality, and the satisfaction value of a specification is a number in the interval $[0, 1]$, which describes the quality of satisfaction. While the use of many-valued semantics in the context of quality aligns with that of robustness, there are notable differences. In particular, our notion of robustness or quality is intrinsic to the logic, while the approach in~\cite{almagor2016quality} requires designers to provide their own interpretation through sets $\mathcal{F}$ or $\mathcal{D}$ of functions. Moreover, there are several choices for defining logical connectives on the interval $[0,1]$, and the suitability of G\"odel's conjunction used in~\cite{almagor2016quality} for formalizing quality is not explicitly addressed. In contrast, we meticulously discuss and motivate all the choices made when defining our semantics with robustness considerations.

Finally, as our work is based on the original works on rLTL~\cite{TabuadaN16}, it is worth mentioning that rLTL has spawned numerous follow-up works, including rLTL model checking~\cite{DBLP:conf/hybrid/AnevlavisNPT19,DBLP:conf/cdc/AnevlavisPNT18,DBLP:journals/tocl/AnevlavisPNT22}, rLTL runtime monitoring~\cite{DBLP:conf/hybrid/MascleNSTW020,Mascle_2022}, and rLTL synthesis~\cite{Nayak22}.
Moreover, several follow-up works have introduced robust extensions of other classes of temporal logics, e.g., Prompt-LTL and Linear Dynamic Logic~\cite{NEIDER2021104810,DBLP:journals/iandc/NeiderWZ22}, Probabilistic Temporal Logics~\cite{Zimmermann23}, and Alternating-Time Temporal Logic~\cite{murano2023robust}.
}

\section{Conclusion}

Inspired by robust LTL, we first developed robust extensions of the logics CTL and CTL*, named rCTL and rCTL*, respectively. Second, we showed that rCTL is more expressive than CTL, while rCTL* is as expressive as CTL*. Third, we showed that the rCTL and rCTL* model checking problem are in $\mathrm{PTIME}$ and $\mathrm{PSPACE}$-complete, respectively, as are the CTL and CTL* model checking problem.
Similarly, we proved that rCTL satisfiability and synthesis are $\mathrm{EXPTIME}$-complete (as are the corresponding problems for CTL) and that rCTL* satisfiability and synthesis are $\mathrm{2EXPTIME}$-complete (as are the corresponding problems for CTL*). So, robustness for branching-time logics does truly come for free.

Tabuada and Neider~\cite{TabuadaN16} described \textit{quality} as the dual of robustness. To illustrate this point, consider the CTL formula $\Diamond \Phi \Rightarrow \Diamond \Psi$. 
According to the motto ``more is better'' we would prefer the system to guarantee the stronger property $\Box\Diamond \Psi$ whenever the environment satisfies the stronger property $\Box\Diamond \Psi$. And similarly, $\Diamond\Box \Phi$ should lead to $\Diamond\Box \Psi$ and $\Box \Phi$ should lead to $\Box \Psi$. Then, a natural question that arises for further research is whether there is an extension of CTL (and CTL*) that can be used to reason about both robustness and quality.



\section*{Statements and Declarations}


\subsection*{Competing interests}
The authors have no competing interests to declare that are relevant to the content of this article.

\subsection*{Funding}
The work was partly funded by the Deutsche Forschungsgemeinschaft (DFG, German Research Foundation) grant number 434592664, by Villum Investigator Grant S4OS, and by the Danish National Research Center DIREC.

\subsection*{Ethics approval}
Not applicable

\subsection*{Consent to participate}
Not applicable

\subsection*{Consent for publication}
Not applicable

\subsection*{Availability of data and materials}
Not applicable

\subsection*{Code availability}
Not applicable

\subsection*{Authors' contributions}
Not applicable





\bibliography{bib}

\end{document}